\documentclass[11pt]{article}

\usepackage{booktabs}  %
\usepackage{multirow}
\usepackage{eepic,epic}
\usepackage{epsfig}
\usepackage{graphicx}
\usepackage{graphics}
\usepackage{color}
\usepackage{ifthen}
\usepackage{mathptmx}
\usepackage{amssymb}
\usepackage[centertags]{amsmath}
\usepackage{pifont}
\usepackage{subfigure} 
\usepackage{paralist}
\usepackage[latin1]{inputenc} 
\usepackage{array,xspace,tikz}
\usetikzlibrary{calc}
\usetikzlibrary{positioning}



\newcommand{\OMIT}[1]{} %

\newenvironment{proofs}{\noindent{\bf Proof.}\hspace*{1em}}{\literalqed\bigskip}

\newenvironment{proofsketch}{\noindent{\bf Proof Sketch.}\hspace*{1em}}{\literalqed\bigskip}

\def\literalqed{{\ \nolinebreak\hfill\mbox{\qedblob\quad}}}

\newcommand{\condition}{\,\mid \:}

\usepackage{ifthen}
\newcommand{\hugeDebug}{false}
\ifthenelse{\equal{\hugeDebug}{false}}{
\newcommand{\normalspacing}{\singlespacing}
}
{
\newcommand{\normalspacing}{\niceninespacing}
}
\ifthenelse{\equal{\hugeDebug}{false}}{
\setlength{\oddsidemargin}{0.0in}
\setlength{\evensidemargin}{\oddsidemargin}
\setlength{\textwidth}{6.7in}
\setlength{\textheight}{8.5in}
\setlength{\topmargin}{-0.0in}
}
{
\setlength{\oddsidemargin}{-0.4in}
\setlength{\evensidemargin}{\oddsidemargin}
\setlength{\textwidth}{5.3in}
}

\ifthenelse{\equal{\hugeDebug}{false}}{
\pagestyle{plain}
}
{
\pagestyle{headings}
}

\makeatletter
\newcommand{\singlespacing}{\let\CS=
\@currsize\renewcommand{\baselinestretch}{1}\tiny\CS}
\newcommand{\singlespacingplus}{\let\CS=
\@currsize\renewcommand{\baselinestretch}{1.25}\tiny\CS}
\newcommand{\doublespacing}{\let\CS=
\@currsize\renewcommand{\baselinestretch}{1.75}\tiny\CS}
\newcommand{\extradoublespacing}{\let\CS=
\@currsize\renewcommand{\baselinestretch}{1.9}\tiny\CS}
\newcommand{\draftspacing}{\let\CS=
\@currsize\renewcommand{\baselinestretch}{2.0}\tiny\CS}
\newcommand{\hugedraftspacing}{\let\CS=
\@currsize\renewcommand{\baselinestretch}{2.4}\tiny\CS}
\newcommand{\niceonespacing}{\let\CS=\@currsize\renewcommand{\baselinestretch}{1.1}\tiny\CS}
\newcommand{\nicetwospacing}{\let\CS=\@currsize\renewcommand{\baselinestretch}{1.2}\tiny\CS}
\newcommand{\nicethreespacing}{\let\CS=\@currsize\renewcommand{\baselinestretch}{1.3}\tiny\CS}
\newcommand{\singlespacingplusplus}{\let\CS=\@currsize\renewcommand{\baselinestretch}{1.35}\tiny\CS}
\newcommand{\nicefourspacing}{\let\CS=\@currsize\renewcommand{\baselinestretch}{1.4}\tiny\CS}
\newcommand{\nicefivespacing}{\let\CS=\@currsize\renewcommand{\baselinestretch}{1.5}\tiny\CS}
\newcommand{\nicesixspacing}{\let\CS=\@currsize\renewcommand{\baselinestretch}{1.6}\tiny\CS}
\newcommand{\nicesevenspacing}{\let\CS=\@currsize\renewcommand{\baselinestretch}{1.7}\tiny\CS}
\newcommand{\niceeightspacing}{\let\CS=\@currsize\renewcommand{\baselinestretch}{1.8}\tiny\CS}
\newcommand{\niceninespacing}{\let\CS=\@currsize\renewcommand{\baselinestretch}{1.9}\tiny\CS}
\makeatother
\normalspacing

\makeatletter \@beginparpenalty=10000 \makeatother

\mathcode`\0="0030      %
\mathcode`\1="0031
\mathcode`\2="0032
\mathcode`\3="0033
\mathcode`\4="0034
\mathcode`\5="0035
\mathcode`\6="0036
\mathcode`\7="0037
\mathcode`\8="0038
\mathcode`\9="0039

\newcount\hour  \newcount\minutes  \hour=\time  \divide\hour by 60
\minutes=\hour  \multiply\minutes by -60  \advance\minutes by \time
\def\mmmddyyyy{\ifcase\month\or Jan\or Feb\or Mar\or Apr\or May\or Jun\or Jul\or
  Aug\or Sep\or Oct\or Nov\or Dec\fi \space\number\day, \number\year}
\def\hhmm{\ifnum\hour<10 0\fi\number\hour :%
  \ifnum\minutes<10 0\fi\number\minutes}
\makeatletter
\def\@cite#1#2{[#1\if@tempswa , #2\fi]}
\makeatother

\makeatletter
\def\@citex[#1]#2{\if@filesw\immediate\write\@auxout{\string\citation{#2}}\fi
  \def\@citea{}\@cite{\@for\@citeb:=#2\do
    {\@citea\def\@citea{,\linebreak[0]}\@ifundefined
       {b@\@citeb}{{\bf ?}\@warning
       {Citation `\@citeb' on page \thepage \space undefined}}%
\hbox{\csname b@\@citeb\endcsname}}}{#1}}
\makeatother

\makeatletter
\def\@cite#1#2{[#1\if@tempswa , #2\fi]}
\def\@citex[#1]#2{\if@filesw\immediate\write\@auxout{\string\citation{#2}}\fi
  \def\@citea{}\@cite{\@for\@citeb:=#2\do
    {\@citea\def\@citea{,\kern1pt\linebreak[0]}\@ifundefined
       {b@\@citeb}{{\bf ?}\@warning
       {Citation `\@citeb' on page \thepage \space undefined}}%
\hbox{\csname b@\@citeb\endcsname}}}{#1}}
\makeatother

\newcounter{alg}

\newenvironment{casedistinction}{\begin{list}
   {{\bf Case~\arabic{case}:}}
   {\usecounter{case}}}{\end{list}}
\newcounter{case}

\newcounter{subcase}

\newcommand\qedblob{\mbox{\ding{113}}}

\newcommand{\probbf}{\rm}
\newcommand{\sat}{{\probbf SAT}}

\newcommand{\muc}{{\probbf \mbox{\sc{}MC$_{\rm{}u}$}}}
\newcommand{\msuc}{{\probbf \mbox{\sc{}MSC$_{\rm{}u}$}}}
\newcommand{\mdc}{{\probbf \mbox{\sc{}MC$_{\rm{}d}$}}}
\newcommand{\msdc}{{\probbf \mbox{\sc{}MSC$_{\rm{}d}$}}}

\newcommand{\mucmember}{{\probbf \mbox{\sc{}MC$_{\rm{}u}$}\hbox{-}\mbox{\sc{}Member}}}
\newcommand{\msucmember}{{\probbf \mbox{\sc{}MSC$_{\rm{}u}$}\hbox{-}\mbox{\sc{}Member}}}
\newcommand{\mdcmember}{{\probbf \mbox{\sc{}MC$_{\rm{}d}$}\hbox{-}\mbox{\sc{}Member}}}
\newcommand{\msdcmember}{{\probbf \mbox{\sc{}MSC$_{\rm{}d}$}\hbox{-}\mbox{\sc{}Member}}}

\newcommand{\muctest}{{\probbf \mbox{\sc{}MC$_{\rm{}u}$}\hbox{-}\mbox{\sc{}Test}}}
\newcommand{\msuctest}{{\probbf \mbox{\sc{}MSC$_{\rm{}u}$}\hbox{-}\mbox{\sc{}Test}}}
\newcommand{\mdctest}{{\probbf \mbox{\sc{}MC$_{\rm{}d}$}\hbox{-}\mbox{\sc{}Test}}}
\newcommand{\msdctest}{{\probbf \mbox{\sc{}MSC$_{\rm{}d}$}\hbox{-}\mbox{\sc{}Test}}}

\newcommand{\mucmemberall}{{\probbf \mbox{\sc{}MC$_{\rm{}u}$}\hbox{-}\allowbreak\mbox{\sc{}Member}\hbox{-}\allowbreak\mbox{\sc{}All}}}
\newcommand{\msucmemberall}{{\probbf \mbox{\sc{}MSC$_{\rm{}u}$}\hbox{-}\allowbreak\mbox{\sc{}Member}\hbox{-}\allowbreak\mbox{\sc{}All}}}
\newcommand{\mdcmemberall}{{\probbf \mbox{\sc{}MC$_{\rm{}d}$}\hbox{-}\allowbreak\mbox{\sc{}Member}\hbox{-}\allowbreak\mbox{\sc{}All}}}
\newcommand{\msdcmemberall}{{\probbf \mbox{\sc{}MSC$_{\rm{}d}$}\hbox{-}\allowbreak\mbox{\sc{}Member}\hbox{-}\allowbreak\mbox{\sc{}All}}}

\newcommand{\mucexists}{{\probbf \mbox{\sc{}MC$_{\rm{}u}$}\hbox{-}\mbox{\sc{}Size}}}
\newcommand{\msucexists}{{\probbf \mbox{\sc{}MSC$_{\rm{}u}$}\hbox{-}\mbox{\sc{}Size}}}
\newcommand{\mdcexists}{{\probbf \mbox{\sc{}MC$_{\rm{}d}$}\hbox{-}\mbox{\sc{}Size}}}
\newcommand{\msdcexists}{{\probbf \mbox{\sc{}MSC$_{\rm{}d}$}\hbox{-}\mbox{\sc{}Size}}}

\newcommand{\mucunique}{{\probbf \mbox{\sc{}MC$_{\rm{}u}$}\hbox{-}\mbox{\sc{}Unique}}}
\newcommand{\msucunique}{{\probbf \mbox{\sc{}MSC$_{\rm{}u}$}\hbox{-}\mbox{\sc{}Unique}}}
\newcommand{\mdcunique}{{\probbf \mbox{\sc{}MC$_{\rm{}d}$}\hbox{-}\mbox{\sc{}Unique}}}
\newcommand{\msdcunique}{{\probbf \mbox{\sc{}MSC$_{\rm{}d}$}\hbox{-}\mbox{\sc{}Unique}}}

\newcommand{\mucfind}{{\probbf \mbox{\sc{}MC$_{\rm{}u}$}\hbox{-}\mbox{\sc{}Find}}}
\newcommand{\msucfind}{{\probbf \mbox{\sc{}MSC$_{\rm{}u}$}\hbox{-}\mbox{\sc{}Find}}}
\newcommand{\mdcfind}{{\probbf \mbox{\sc{}MC$_{\rm{}d}$}\hbox{-}\mbox{\sc{}Find}}}
\newcommand{\msdcfind}{{\probbf \mbox{\sc{}MSC$_{\rm{}d}$}\hbox{-}\mbox{\sc{}Find}}}

\newtheorem{theorem}{Theorem}[section]

\newtheorem{definition}[theorem]{Definition}
\newtheorem{lemma}[theorem]{Lemma}

\newtheorem{claim}[theorem]{Claim}
\newtheorem{construction}[theorem]{Construction}

\newcommand{\p}{\ensuremath{\mathrm{P}}}
\newcommand{\np}{\ensuremath{\mathrm{NP}}}
\newcommand{\DP}{\ensuremath{\mathrm{DP}}}
\newcommand{\conp}{\ensuremath{\mathrm{coNP}}}

\usepackage{xspace}

\newcommand{\thmref}[1]{Theorem~\ref{#1}}

\newcommand{\consref}[1]{Construction~\ref{#1}}

\newcommand{\ie}{i.e.,\xspace}

\newcommand{\bx}{\overline{x}}
\newcommand{\bu}{\overline{u}}
\newcommand{\ha}{\widehat{a}}

\title{The Complexity of Computing Minimal Unidirectional Covering
  Sets\thanks{
This work was supported in part by
    DFG grants BR-2312/6-1, RO-1202/12-1 (within the European Science
    Foundation's EUROCORES program LogICCC), BR~2312/3-2, and
    RO-1202/11-1, and by the Alexander von Humboldt Foundation's
    TransCoop program.  This work was done in part while the fifth
    author was visiting the University of Rochester.}}

\author{%
Dorothea Baumeister \\ 
Institut f\"ur Informatik \\
Heinrich-Heine-Universit{\"a}t D{\"u}sseldorf \\
40225 D\"usseldorf, Germany
\and
Felix Brandt \\
Institut f{\"u}r Informatik \\
Ludwig-Maximilians-Universit{\"a}t M{\"u}nchen \\
80538 M{\"u}nchen, Germany
\and
Felix Fischer \\
Institut f{\"u}r Informatik \\
Ludwig-Maximilians-Universit{\"a}t M{\"u}nchen \\
80538 M{\"u}nchen, Germany
\and
Jan Hoffmann \\
Institut f{\"u}r Informatik \\
Ludwig-Maximilians-Universit{\"a}t M{\"u}nchen \\
80538 M{\"u}nchen, Germany
\and
J{\"o}rg Rothe \\ 
Institut f\"ur Informatik \\
Heinrich-Heine-Universit{\"a}t D{\"u}sseldorf \\
40225 D\"usseldorf, Germany
}

\date{July 15, 2009}

\begin{document}
\sloppy

\maketitle 
\begin{abstract}
Given a binary dominance relation on a set of alternatives, a common
thread in the social sciences is to identify subsets of alternatives
that satisfy certain  notions of stability. Examples can be found in
areas as diverse as
voting theory, game theory, and argumentation theory. Brandt and
Fischer~\cite{bra-fis:j:minimal-covering-set} proved that it is
$\np$-hard to decide whether an alternative is contained in some
inclusion-minimal unidirectional (i.e., either upward or downward)
covering set.
For both problems, we raise this %
lower bound to the $\Theta_2^p$ level of the polynomial hierarchy and
provide a $\Sigma_2^p$ upper bound.  Relatedly, we show that a variety
of other natural problems regarding minimal or minimum-size 
unidirectional covering sets 
are hard or complete for either of NP, coNP, and~$\Theta_2^p$.  An important
consequence of our results is that neither minimal upward nor minimal
downward covering sets (even when guaranteed to exist) can be computed
in polynomial time unless $\p=\np$.  This sharply contrasts
with Brandt and Fischer's result that minimal bidirectional covering
sets (i.e., sets that are both minimal upward and minimal downward
covering sets) are polynomial-time computable.
\end{abstract}

\section{Introduction}
\label{sec:introduction}

A common thread in the social sciences is to identify sets of
alternatives that satisfy certain notions of stability according to
some binary dominance relation. Applications range from cooperative to
non-cooperative game theory, 
from social choice theory to argumentation theory, and 
from multi-criteria decision analysis to sports tournaments (see, e.g.,
\cite{las:b:tournament-solutions-majority-voting,bra-fis:j:minimal-covering-set}
and the references therein).

In social choice settings, 
the most common dominance relation is the pairwise
majority relation, where an alternative~$x$ is said to dominate another
alternative~$y$ if the number of individuals preferring~$x$ to~$y$
exceeds the number of individuals preferring~$y$
to~$x$.  McGarvey~\cite{mcg:j:election-graph}
proved that \emph{every} asymmetric dominance
relation can be realized via a particular preference profile, even if
the individual preferences are
linear.  For example, Condorcet's well-known paradox says that the
majority relation may contain cycles and thus does not always have
maximal elements, even if all of the underlying individual
preferences do.  This means that the concept of maximality is rendered
useless in many cases, which is why various so-called \emph{solution
concepts} have been proposed.  Solution concepts can be used in place
of maximality for nontransitive relations (see, e.g.,
\cite{las:b:tournament-solutions-majority-voting}).  In particular,
concepts based on so-called \emph{covering relations}---transitive 
subrelations of the
dominance relation at hand---have turned out to be very attractive
\cite{fis:j:condorcet,mil:j:tournaments-and-majority-voting,dut:j:covering-sets}.

Computational social choice is an emerging new field at the interface
of social choice theory, economics, and computer science that focuses
on the computational properties of social-choice-related concepts and
problems~\cite{che-end-lan-mau:c:computational-social-choice}.  For
example, voting procedures---and dominance-based solution concepts are closely related
to the winner-determination problem in certain voting systems---have
applications in artificial intelligence (especially in multiagent
systems), in aggregating the web-page rankings from multiple search
engines (see Dwork et al.~\cite{dwo-kum-nao-siv:c:rank-aggregation}),
and other domains of computer science.  That is why the computational
properties of voting and other social-choice-related notions have been
studied in-depth recently (see the survey~\cite{fal-hem-hem-rot:b:richer}).

This paper studies the computational complexity of problems related to
the notions of upward and downward covering sets in dominance
graphs.  An
alternative~$x$ is said to \emph{upward cover} another alternative $y$
if $x$ dominates $y$ and every alternative dominating $x$ also
dominates~$y$.  The intuition is that $x$ ``strongly'' dominates $y$
in the sense that there is no alternative that dominates $x$ but
not~$y$.  Similarly, an alternative~$x$ is said to \emph{downward
  cover} another alternative $y$ if $x$ dominates $y$ and every
alternative dominated by $y$ is also dominated by~$x$.  The intuition
here is that $x$ ``strongly'' dominates $y$ in the sense that there is
no alternative dominated by $y$ but not by~$x$.  A \emph{minimal
  upward} or \emph{minimal downward covering set} is defined as an
inclusion-minimal set of alternatives that satisfies certain notions
of internal and external stability with respect to the upward or
downward covering
relation~\cite{dut:j:covering-sets,bra-fis:j:minimal-covering-set}.

Recent work in theoretical computer science has addressed
the computational complexity of most solution concepts proposed in the
context of binary dominance (see, e.g.,
\cite{woe:j:banks-winners-in-tournaments,alo:j:ranking-tournaments,con:c:slater-rankings,bra-fis-har:c:complexity-of-choice-sets,bra-fis:j:minimal-covering-set,bra-fis-har-mai:c:tournament-equilibrium-set}).
In particular, Brandt and
Fischer~\cite{bra-fis:j:minimal-covering-set}
have shown $\np$-hardness of both
the problem of deciding whether an alternative is contained in some
minimal upward and the problem of deciding whether an alternative is
contained in some minimal downward covering set, is $\np$-hard.
For both problems, we improve on these results by raising
their $\np$-hardness lower bounds to the $\Theta_2^p$ level
of the polynomial hierarchy, and we provide an upper bound of~$\Sigma_2^p$.
Moreover, we will analyze the complexity of a variety of
other problems associated with minimal and minimum-size
upward and downward covering sets that have not
been studied before.  In particular, we provide hardness and
completeness results for the complexity classes $\np$, $\conp$,
and~$\Theta_2^p$.
Remarkably, these new results imply that neither minimal upward
covering sets nor minimal downward covering sets (even when guaranteed to exist) can be found in
polynomial time unless $\p=\np$.  This sharply contrasts with Brandt
and Fischer's result that minimal \emph{bidirectional} covering sets
(i.e., sets that are both minimal upward and minimal downward covering
sets) are polynomial-time
computable~\cite{bra-fis:j:minimal-covering-set}.  Note that,
notwithstanding the hardness of computing minimal upward covering
sets, the decision version of this search problem is trivially
in~$\p$: Every dominance graph always contains a minimal upward
covering set.

Our $\Theta_2^p$-hardness results apply Wagner's
method~\cite{wag:j:min-max} that was useful also in other contexts
(see, e.g.,
\cite{wag:j:min-max,hem-hem-rot:j:dodgson,hem-rot:j:max-independent-set-by-greed,hem-wec:j:mee,hem-rot-spa:j:vcgreedy}).
To the best of our knowledge, our constructions for the first time
apply his method to problems defined in terms of minimality rather
than minimum size of a solution.

\section{Definitions and Notation}
\label{sec:prelims}

In this section, we define the required notions and notation from
social choice theory and complexity theory.

\begin{definition}[Covering Relations]
Let $A$ be a finite set of alternatives, 
let $B\subseteq A$, and
let $\succ\ \subseteq
A{\times}A$ be a dominance relation on~$A$, i.e., $\succ$ is
asymmetric and irreflexive.\footnote{In general, $\succ$ need not be
transitive or complete.  For alternatives $x$ and $y$, $x \succ y$
(equivalently, $(x,y) \in\; \succ$) is interpreted as $x$ being strictly
preferred to $y$ (and we say ``$x$ dominates $y$''), for example as
the result of a strict majority of voters preferring $x$ to~$y$.}
A dominance relation $\succ$ on a set $A$ of alternatives can be
conveniently
represented as a \emph{dominance graph}, denoted by $(A,\succ)$, whose
vertices are the alternatives from~$A$, and for each $x, y \in A$
there is a directed edge
from $x$ to $y$ if and only if $x \succ y$.

For any two alternatives $x$ and $y$ in~$B$, define the following
covering relations (see, e.g., \cite{fis:j:condorcet,mil:j:tournaments-and-majority-voting,bor:j:consistent-majoritarian-choice}):
\begin{itemize}
\item 
\emph{$x$ upward covers $y$ in $B$}, denoted by $x\, C_u^B\, y$, if $x \succ
y$ and for all $z \in B$, $z \succ x$ implies $z \succ y$, and
\item \emph{$x$ downward covers $y$ in $B$}, denoted by $x\, C_d^B\, y$, if $x
\succ y$ and for all $z \in B$, $y \succ z$ implies $x \succ z$.

\end{itemize}
When clear from the context, we omit mentioning ``in $B$'' explicitly
and simply write $x\, C_u\, y$ rather than $x\, C_u^B\, y$, and
$x\, C_d\, y$ rather than $x\, C_d^B\, y$. 

\end{definition}

\begin{definition}[Uncovered Set] 
Let $A$ be a set of alternatives, let $B\subseteq A$ be any subset,
let $\succ$ be a dominance relation on~$A$, and let $C$ be a covering
relation on $A$ based on~$\succ$. The \emph{uncovered set of $B$ with
respect to $C$} is defined as
\[
\mathrm{UC}_C(B)=\{x\in B \condition y\, C\, x \mbox{ for no } y \in
B\}.
\]
For notational convenience, let $\mathrm{UC}_x(B) =
\mathrm{UC}_{C_x}(B)$ for
$x \in \{u,d\}$, and we call
$\mathrm{UC}_u(B)$ the \emph{upward uncovered set of $B$} and
$\mathrm{UC}_d(B)$ the \emph{downward uncovered set of~$B$}.
\end{definition}

For
both 
the upward 
and the downward 
covering relation (henceforth unidirectional covering relations), transitivity of the relation implies nonemptiness
of the corresponding uncovered set for each nonempty set of alternatives.
Every 
upward uncovered set contains one or more minimal 
upward covering sets, whereas minimal downward covering sets may not
always exist~\cite{bra-fis:j:minimal-covering-set}.
Dutta~\cite{dut:j:covering-sets} proposed minimal covering sets in the
context of tournaments, i.e., complete dominance relations, where both notions of covering coincide. Minimal
unidirectional covering sets are one of several possible generalizations to
incomplete dominance relations (for more details, see
\cite{bra-fis:j:minimal-covering-set}).  The intuition underlying
covering sets is that there should be no reason to restrict the
selection by excluding some alternative from it (internal stability)
and there should be an argument against each proposal to include an
outside alternative into the selection (external stability).

\begin{definition}[Minimal Covering Set]
Let $A$ be a set of alternatives,
let $\succ$ be a dominance relation on~$A$,
and let $C$ be a covering relation based on~$\succ$.
A subset
$B \subseteq A$ is a \emph{covering set for $A$
 under $C$}
if the following two properties hold:
\begin{itemize}
\item \emph{Internal stability:} $\mathrm{UC}_C(B)=B$.
\item \emph{External stability:} For all $x\in A - B$, $x \not\in
\mathrm{UC}_C(B \cup \{x\})$.
\end{itemize}

A covering set $M$ for $A$ under $C$
is said to be \text{\emph{(inclusion-)minimal}} if no $M' \subset M$
is 
a covering set for~$A$
 under~$C$.
\end{definition}

Occasionally, it might be helpful to specify the dominance relation
explicitly to avoid ambiguity.  In such cases we refer to the
dominance graph used and write, e.g., ``$M$ is an upward covering set
for~$(A,\succ)$.''

In addition to the (inclusion-)minimal 
unidirectional
covering sets considered
in~\cite{bra-fis:j:minimal-covering-set}, we will also consider
\emph{minimum-size} 
covering sets, i.e., unidirectional
covering sets of smallest cardinality.  For some of the computational
problems we study, different complexities can be shown for the
minimal and minimum-size versions of the problem
(see Theorem~\ref{thm:results} and Table~\ref{tab:results}).
Specifically,
we will consider six types of computational problems, for both upward
and downward covering sets, and for each both their ``minimal'' and
``minimum-size'' versions.  We first define the six problem types for
the case of minimal upward covering sets:
\begin{enumerate}
\item $\mucexists$: Given a set $A$ of alternatives, a dominance
  relation $\succ$ on~$A$, and a positive integer~$k$, does there
  exist some minimal upward covering set for $A$ containing at most
  $k$ alternatives?

\item $\mucmember$: Given a set $A$ of alternatives, a dominance
  relation $\succ$ on~$A$, and a distinguished element $d \in A$, is
  $d$ contained in some minimal upward covering set for~$A$?

\item $\mucmemberall$: Given a set $A$ of alternatives, a dominance
  relation $\succ$ on~$A$, and a distinguished element $d \in A$, is
  $d$ contained in all minimal upward covering sets for~$A$?

\item $\mucunique$: Given a set $A$ of alternatives and a dominance
  relation $\succ$ on~$A$, does there exist a unique minimal upward
  covering set for~$A$?

\item $\muctest$: Given a set $A$ of alternatives, a dominance relation
  $\succ$ on~$A$, and a subset $M \subseteq A$, is $M$ a minimal
  upward covering set for~$A$?

\item $\mucfind$: Given a set $A$ of alternatives and a dominance
  relation $\succ$ on~$A$, find a minimal upward covering set for~$A$.
\end{enumerate}

If we replace ``upward'' by ``downward'' above, we obtain the six
corresponding ``downward covering'' versions, denoted by $\mdcexists$,
$\mdcmember$, $\mdcmemberall$, $\mdcunique$, $\mdctest$, and $\mdcfind$.
And if we replace ``minimal'' by ``minimum-size'' in the twelve
problems just defined, we obtain the corresponding ``minimum-size''
versions: $\msucexists$, $\msucmember$, $\msucmemberall$,
$\msucunique$, $\msuctest$, $\msucfind$, $\msdcexists$, $\msdcmember$,
$\msdcmemberall$, $\msdcunique$, $\msdctest$, and $\msdcfind$.

Note that the four problems $\mucfind$, $\mdcfind$, $\msucfind$, and
$\msdcfind$ are search problems, whereas the other twenty problems are
decision problems.

We assume that the reader is familiar with the basic notions of
complexity theory, such as polynomial-time many-one reducibility
and the related notions of hardness and completeness, and also with
standard complexity classes such as $\p$, $\np$, $\conp$,
and the polynomial hierarchy~\cite{mey-sto:b:reg-exp-needs-exp-space}
(see also, e.g., the textbooks
\cite{pap:b-1994:complexity,rot:b:cryptocomplexity}).  In particular,
$\conp$ is the class of sets whose complements are in~$\np$.
$\Sigma_2^p = \np^{\np}$, the second level of the polynomial hierarchy,
consists of all sets that can be solved by an $\np$ oracle machine that has
access (in the sense of a Turing reduction) to an $\np$ oracle set such as
$\sat$.  $\sat$ denotes the satisfiability problem of propositional logic,
which is one of the standard $\np$-complete problems (see, e.g., Garey
and Johnson~\cite{gar-joh:b:int}) and is defined as follows: Given a
boolean formula in conjunctive normal form, does there exist a truth
assignment to its variables that satisfies the formula?

Papadimitriou and Zachos~\cite{pap-zac:c:two-remarks} introduced
the class of problems that can be decided by a $\p$
machine that accesses its $\np$ oracle in a parallel manner.  This
class is also known as the $\Theta_2^p$ level of the polynomial hierarchy
(see Wagner~\cite{wag:j:bounded}), and has been shown to coincide with
the class of problems solvable in polynomial time via asking
$\mathcal{O}(\log n)$ sequential Turing queries to~$\np$ (see
\cite{hem:c:sky-stoc,koe-sch-wag:j:diff}).  Equivalently,
$\Theta_2^p$ is the closure of $\np$ under polynomial-time
truth-table reductions.  It follows immediately from the definitions
that $\p \subseteq \np \cap \conp \subseteq \np \cup \conp \subseteq
\Theta_2^p \subseteq \Sigma_2^p$.

$\Theta_2^p$ captures the complexity of various optimization
problems.  For example, the problem of testing whether the size of a
maximum clique in a given graph is an odd number, the problem of
deciding whether two given graphs have minimum vertex covers of the
same size, and the problem of recognizing those graphs for which
certain heuristics yield good approximations for the size of a maximum
independent set or for the size of a minimum vertex cover each are
known to be complete for $\Theta_2^p$ (see
\cite{wag:j:min-max,hem-rot:j:max-independent-set-by-greed,hem-rot-spa:j:vcgreedy}).
Hemaspaandra and Wechsung~\cite{hem-wec:j:mee} proved that the
minimization problem for boolean formulas is $\Theta_2^p$-hard.  In
the field of computational social choice, the winner problems for
Dodgson~\cite{dod:unpubMAYBE:dodgson-voting-system},
Young~\cite{you:j:extending-condorcet}, and 
Kemeny~\cite{kem:j:mathematics-without-numbers} elections have been shown
to be $\Theta_2^p$-complete in the nonunique-winner model
\cite{hem-hem-rot:j:dodgson,rot-spa-vog:j:young,hem-spa-vog:j:kemeny},
and also in the unique-winner
model~\cite{hem-hem-rot:tCORR-RevisedWithConfPtr:hybrid}.

\section{Results and Discussion}
\label{sec:results}

\subparagraph{Results.}
Brandt and Fischer~\cite{bra-fis:j:minimal-covering-set} proved that
it is $\np$-hard to decide whether a given alternative is contained in
some minimal unidirectional covering
set.  Using the notation of this paper, their results state that the problems
$\mucmember$ and $\mdcmember$ are $\np$-hard.  The question of whether
these two problems are $\np$-complete or of higher complexity was
left open in~\cite{bra-fis:j:minimal-covering-set}.  Our contribution is
\begin{enumerate}
\item to raise Brandt and Fischer's $\np$-hardness lower bounds for
  $\mucmember$ and $\mdcmember$ to $\Theta_2^p$-hardness and to
  provide (simple) $\Sigma_2^p$ upper bounds for these problems, and
\item to extend the techniques we developed to apply also to the 22
  other covering set problems
defined in Section~\ref{sec:prelims}, in particular to the search
problems.
\end{enumerate}
Our results are stated in the following theorem.

\begin{theorem}
\label{thm:results}
The complexity of the covering set problems defined in
Section~\ref{sec:prelims} is as shown in Table~\ref{tab:results}.
\end{theorem}

\begin{table}
\centering
\footnotesize
\begin{tabular}{lllll}
\toprule
Problem Type
 & \multicolumn{1}{c}{$\muc$}
 & \multicolumn{1}{c}{$\msuc$}
 & \multicolumn{1}{c}{$\mdc$}
 & \multicolumn{1}{c}{$\msdc$} \\
\midrule
{\sc Size}
 & $\np$-complete
 & $\np$-complete
 & $\np$-complete
 & $\np$-complete \\[0.5ex]
{\sc Member}
 & $\Theta_2^p$-hard and in $\Sigma_2^p$
 & $\Theta_2^p$-complete
 & $\Theta_2^p$-hard and in $\Sigma_2^p$
 & $\conp$-hard and in $\Theta_2^p$ \\[0.5ex]
{\sc Member-All}
 & $\conp$-complete~\cite{bra-fis:j:minimal-covering-set}
 & $\Theta_2^p$-complete
 & $\conp$-complete~\cite{bra-fis:j:minimal-covering-set}
 & $\conp$-hard and in $\Theta_2^p$ \\[0.5ex]
{\sc Unique}
 & $\conp$-hard and in $\Sigma_2^p$
 & $\conp$-hard and in $\Theta_2^p$
 & $\conp$-hard and in $\Sigma_2^p$
 & $\conp$-hard and in $\Theta_2^p$ \\[0.5ex]
{\sc Test}
 & $\conp$-complete
 & $\conp$-complete
 & $\conp$-complete
 & $\conp$-complete \\[0.5ex]
{\sc Find}
 & not in polynomial
 & not in polynomial
 & not in polynomial
 & not in polynomial \\
 & time unless $\p = \np$
 & time unless $\p = \np$
 & time unless $\p = \np$
 &  time unless $\p = \np$ \\
 & & & (follows from \cite{bra-fis:j:minimal-covering-set}) & \\
\bottomrule
\end{tabular}
\caption{Overview of complexity results for the various types of
  covering set problems.  As indicated, previously known results are due to
  Brandt and Fischer~\cite{bra-fis:j:minimal-covering-set}; all other
  results are new to this paper.}
\label{tab:results}
\end{table}

The detailed proofs of the single results collected in
Theorem~\ref{thm:results}
will be presented in Section~\ref{sec:proofs}, and the technical
constructions establishing the properties that are needed for these
proofs are given in Section~\ref{sec:constructions}.

\subparagraph{Discussion.}
We consider the problems of \emph{finding} minimal and minimum-size upward
and downward covering sets ($\mucfind$, $\mdcfind$, $\msucfind$, and
$\msdcfind$) to be particularly important and natural.

Regarding upward covering sets, we stress that our result (see
Theorem~\ref{thm:search}) that, assuming $\p \neq \np$, $\mucfind$ and
$\msucfind$ are hard to compute does not follow directly from the
$\np$-hardness of $\mucmember$ in any obvious way.\footnote{The
  decision version of
  $\mucfind$ is: Given a dominance graph, does it contain a
  minimal upward covering set?  However, this question has always an
  affirmative answer, so the decision version of $\mucfind$ is
  trivially in~$\p$.  Note also that $\mucfind$ can be reduced in a
  ``disjunctive truth-table'' fashion to the search version of $\mucmember$
  (``Given a dominance graph $(A, \succ)$ and an alternative $d \in A$,
  find some minimal upward covering set for~$A$ that contains $d$'')
  by asking this oracle set about all alternatives in parallel.
  So $\mucfind$ is no harder (with respect to disjunctive truth-table
  reductions) than that problem.  The converse,
  however, is not at all obvious. Brandt and Fischer's results only
  imply the hardness of finding an alternative that is contained in
  \emph{all} minimal upward covering sets~\cite{bra-fis:j:minimal-covering-set}.}
Our reduction that raises the lower bound of $\mucmember$ from
$\np$-hardness to $\Theta_2^p$-hardness, however, also allows us to
prove that $\mucfind$ and $\msucfind$ cannot be solved in polynomial
time unless $\p = \np$.

Regarding downward covering sets, that $\mdcfind$ cannot be computed
in polynomial time unless $\p = \np$ is an immediate consequence of
Brandt and Fischer's result that it is $\np$-complete to decide
whether there exists a minimal downward covering
set~\cite[Thm.~9]{bra-fis:j:minimal-covering-set}.  We provide as
Theorem~\ref{thm:down-search} an alternative proof based on our
reduction showing that $\mdcmember$ is $\Theta_2^p$-hard.  In contrast to Brandt and Fischer's proof, our proof shows the hardness of $\mdcfind$ even when the existence of a (minimal) downward covering set is guaranteed.
As
indicated in Table~\ref{tab:results}, $\conp$-completeness of
$\mucmemberall$ and $\mdcmemberall$ was also shown previously by
Brandt and Fischer~\cite{bra-fis:j:minimal-covering-set}.

As mentioned above, the two problems $\mucmember$ and $\mdcmember$
were already known to be
$\np$-hard~\cite{bra-fis:j:minimal-covering-set} and are here shown to
be even $\Theta_2^p$-hard.  One may naturally wonder whether raising
their (or any problem's) lower bound from $\np$-hardness to
$\Theta_2^p$-hardness gives us any more insight into the problem's
inherent computational complexity.  After all, $\p = \np$ if and only
if $\p = \Theta_2^p$.  However, this question is a bit more subtle
than that and has been discussed carefully by Hemaspaandra et
al.~\cite{hem-hem-rot:j:raising-lower-bounds-survey}.  They make the
case that the answer to this question crucially depends on what one
considers to be the most natural computational model.  In particular,
they argue that raising $\np$-hardness to $\Theta_2^p$-hardness
potentially (i.e., unless longstanding open problems regarding the
separation of the corresponding complexity classes could be solved) is
an improvement in terms of randomized polynomial time and in terms of
unambiguous polynomial
time~\cite{hem-hem-rot:j:raising-lower-bounds-survey}.

\section{Constructions}
\label{sec:constructions}

In this section, we provide the constructions that will be used in
Section~\ref{sec:proofs} to obtain the new complexity results for the
problems
defined in Section~\ref{sec:prelims}.

\subsection{Minimal and Minimum-Size Upward Covering Sets}

We start by giving the constructions that will be used for establishing
results on the minimal and minimum-size upward covering set problems.
Brandt and Fischer~\cite{bra-fis:j:minimal-covering-set} proved the
following result.  Since we will need their reduction in 
Construction~\ref{cons:theta} and Section~\ref{sec:proofs},
we give a proof sketch for Theorem~\ref{thm:up_np}.

\begin{theorem}[Brandt and Fischer~\cite{bra-fis:j:minimal-covering-set}] 
\label{thm:up_np}
Deciding whether a designated alternative is contained in some minimal
upward covering set for a given dominance graph 
is $\np$-hard.  That is, $\mucmember$ is $\np$-hard.
\end{theorem}

\begin{proofsketch} 
$\np$-hardness is shown by a reduction from $\sat$.  Given a boolean
formula in conjunctive normal form,
$\varphi(v_1, v_2, \dots , v_n) = c_1 \wedge c_2 \wedge \dots 
\wedge c_r$, over the set $V = \{v_1, v_2, \dots , v_n\}$ of variables, 
construct an instance $(A,\succ,d)$ of
$\mucmember$ 
as follows.  The set of
alternatives is
\[
A=\{x_i,\overline{x}_i,x_i',\overline{x}_i' \condition v_i \in V\}
\cup \{y_j \condition c_j \mbox{ is a clause in } \varphi\} \cup
\{d\},
\]
where $d$ is the distinguished alternative whose membership in a
minimal upward covering set for $A$ is to be decided, and the
dominance relation $\succ$ is defined by:
\begin{itemize}
\item For each~$i$, $1 \leq i \leq n$, there is a cycle $x_i \succ
\overline{x}_i \succ x_i' \succ \overline{x}_i' \succ x_i$;
\item if variable $v_i$ occurs in clause $c_j$ as a
positive literal, then $x_i \succ y_j$;
\item if variable $v_i$ occurs in clause
$c_j$ as a negative literal, then $\overline{x}_i \succ y_j$; and 
\item for each~$j$, $1 \leq j \leq r$, we have $y_j \succ d$.
\end{itemize}

\begin{figure}[tb]
  \centering
	\begin{tikzpicture}[scale=1]
  \tikzstyle{every node}=[circle,draw,minimum size=7mm,inner sep=0pt,font=\scriptsize]
	\draw (0,0) node(x14){$\bx_1'$} ++(0,1.5) node(x11){$x_1$} ++(1.5,0) node(x12){$\bx_1$} ++(0,-1.5) node(x13){$x_1'$} ++(2,0) node(x24){$\bx_2'$} ++(0,1.5) node(x21){$x_2$} ++(1.5,0) node(x22){$\bx_2$} ++(0,-1.5) node(x23){$x_2'$} ++(2,0) node(x34){$\bx_3'$} ++(0,1.5) node(x31){$x_3$} ++(1.5,0) node(x32){$\bx_3$} ++(0,-1.5) node(x33){$x_3'$};
	\foreach \i in {1,2,3} {
		\foreach \x / \y in {x\i1/x\i2,x\i2/x\i3,x\i3/x\i4,x\i4/x\i1}
			{ \draw[-latex] (\x) -- (\y); }}
	\path (x12) -- node[above=1.5cm](y1){$y_1$} (x21);
	\path (x22) -- node[above=1.5cm](y2){$y_2$} (x31);
	\path (y1) -- node[draw,above=.5cm](d){$d$} (y2);
	\foreach \x / \y in {x11/y1,x12/y2,x22/y1,x31/y1,x32/y2,y1/d,y2/d}
		{ \draw[-latex] (\x) -- (\y); }
\end{tikzpicture}
	\caption{Dominance graph for \thmref{thm:up_np}, example for the formula $(v_1 \vee \neg v_2 \vee v_3) \wedge (\neg v_1 \vee \neg v_3)$.}
  \label{fig:up_np}
\end{figure}

As an example of this reduction, Figure~\ref{fig:up_np} shows the
dominance graph resulting from the formula $(v_1 \vee \neg v_2 \vee v_3) \wedge (\neg v_1 \vee \neg v_3)$, which is
satisfiable, for example via the truth assignment that sets each of
$v_1$, $v_2$, and $v_3$ to false.  Note that in this case the set
$\{\overline{x}_1, \overline{x}'_1,
   \overline{x}_2, \overline{x}'_2,
   \overline{x}_3, \overline{x}'_3\} \cup \{d\}$
is a minimal upward covering set for~$A$, so there indeed exists a
minimal upward covering set for $A$ that
contains the designated alternative~$d$.
In general, Brandt and
Fischer~\cite{bra-fis:j:minimal-covering-set} proved that there exists
a satisfying assignment for $\varphi$ if and only if $d$ is contained
in some minimal upward covering set for~$A$.~\end{proofsketch}

As we will use this reduction
to prove results for both $\mucmember$ and 
some of the other problems stated in 
Section~\ref{sec:prelims}, we now analyze the minimal and
minimum-size upward covering sets of the dominance graph
constructed in the proof sketch of Theorem~\ref{thm:up_np}.
Brandt and Fischer~\cite{bra-fis:j:minimal-covering-set} showed that
each minimal upward covering set for $A$ contains exactly two of the four
alternatives corresponding to any of the variables, i.e., either $x_i$ and
$x_i'$, or $\overline{x_i}$ and $\overline{x_i}'$, $1 \leq i \leq n$.
We now assume that if $\varphi$ is not satisfiable then for each truth
assignment to the variables of~$\varphi$,
at least two clauses are unsatisfied (which can be ensured, if needed,
by adding two dummy variables). It follows that every minimal upward covering set
for $A$ not containing alternative $d$ must consist of at least $2n+2$
alternatives, and
every minimal upward covering set for $A$ containing $d$ consists of
exactly $2n+1$ alternatives.
Thus, $\varphi$ is satisfiable if and only if every minimum-size upward
covering set consists of $2n+1$ alternatives and contains $d$.

We now provide another construction that transforms a given
boolean formula into a dominance graph with quite different properties.
\begin{construction}[To be used for showing coNP-hardness for upward 
covering set problems]~
\label{cons:conp}
Given a boolean formula in conjunctive normal form,
$\varphi(w_1, w_2, \dots, w_k)=f_1 \wedge f_2
\wedge \dots \wedge f_{\ell}$, over the set $W = \{w_1, w_2, \dots,
w_k\}$ of variables, we construct a set of alternatives $A$ and a
dominance relation $\succ$ on~$A$.

The set of alternatives is
$A=\{u_i,\overline{u}_i,u_i',\overline{u}_i' \condition w_i \in W\}
\cup \{e_j, e_j' \condition f_j \mbox{ is a clause in } \varphi\}
\cup \{a_1,a_2,a_3\}$,
and the dominance relation $\succ$ is
defined by:
\begin{itemize}
\item
For each~$i$, $1\leq i \leq k$, there is a cycle $u_i \succ
\overline{u}_i \succ u_i' \succ \overline{u}_i' \succ u_i$;
\item if variable $w_i$ occurs in clause $f_j$ as a positive literal,
then $u_i \succ e_j$, $u_i \succ e'_j$, 
$e_j \succ \overline{u}_i$, and $e'_j \succ \overline{u}_i$;
\item if variable $w_i$ occurs in clause $f_j$ as a negative literal,
then $\overline{u}_i \succ e_j$, $\overline{u}_i \succ e'_j$,
$e_j \succ u_i$, and $e'_j \succ u_i$;
\item if variable $w_i$ does not occur in clause~$f_j$, 
then $e_j \succ u_i'$ and $e_j' \succ \overline{u}_i'$;
\item for each $j$, $1\leq j \leq \ell$, we have $a_1 \succ e_j$ and
$a_1 \succ e'_j$; and
\item there is a cycle $a_1 \succ a_2 \succ a_3 \succ a_1$.
\end{itemize}
\end{construction}

Figure~\ref{fig:up_gadgets} shows some parts of the dominance
graph that results from the given boolean formula~$\varphi$.  In
particular, Figure~\ref{fig:up_pos} shows that part of this graph
that corresponds to some variable $w_i$ occuring in clause $f_j$ as a
positive literal; Figure~\ref{fig:up_neg} shows that part of this
graph that corresponds to some variable $w_i$ occuring in clause $f_j$
as a negative literal; and Figure~\ref{fig:up_nil} shows that
part of this graph that corresponds to some variable $w_i$ not
occuring in clause~$f_j$.

\begin{figure}[tb]
  \centering
	\begin{tikzpicture}[scale=1]
  \tikzstyle{every node}=[circle,draw,minimum size=7mm,inner sep=0pt,font=\scriptsize]
	\draw (0,0) node(x14){$\bu_1'$} ++(0,1.5) node(x11){$u_1$} ++(1.5,0) node(x12){$\bu_1$} ++(0,-1.5) node(x13){$u_1'$} ++(2,0) node(x24){$\bu_2'$} ++(0,1.5) node(x21){$u_2$} ++(1.5,0) node(x22){$\bu_2$} ++(0,-1.5) node(x23){$u_2'$} ++(2,0) node(x34){$\bu_3'$} ++(0,1.5) node(x31){$u_3$} ++(1.5,0) node(x32){$\bu_3$} ++(0,-1.5) node(x33){$u_3'$};
	\foreach \i in {1,2,3} {
		\foreach \x / \y in {x\i1/x\i2,x\i2/x\i3,x\i3/x\i4,x\i4/x\i1}
			{ \draw[-latex] (\x) -- (\y); }}
	\draw (x12) ++(60:2cm) +(0,.5) node(e1){$e_1$} (x22) ++(60:2cm) +(0,.5) node(e3){$e_1'$} (x13) ++(300:2cm) +(0,-.5) node(e2){$e_2$} (x23) ++(300:2cm) +(0,-.5) node(e4){$e_2'$};
	\path (e1) -- node[draw,above=1cm](a1){$a_1$} (e3);
	\foreach \x / \y in {e1/x11,x12/e1,x12/e3,x21/e1,e1/x22,x21/e3,e3/x22}
		{ \draw[-latex] (\x) -- (\y); }
	\draw[-latex] (e1) .. controls +(350:7cm) and +(60:3cm) .. (x33);
	\draw[-latex] (e3) .. controls +(down:1.5cm) and +(120:1cm) .. (x34);
	\draw[-latex] (e3) .. controls +(195:1cm) and +(20:1cm) .. (x11);
	\foreach \x / \y in {e2/x23,e4/x24}
		{ \draw[-latex] (\x) -- (\y); }
	\draw[-latex] (x32) .. controls +(300:3cm) and +(10:7cm) .. (e2);
	\draw[-latex] (x11) .. controls +(240:3cm) and +(170:7cm) .. (e4);
	\draw[-latex] (e2) .. controls +(27:3cm) and +(230:2cm) .. (x31);
	\draw[-latex] (e4) .. controls +(153:2cm) and +(310:3cm) .. (x12);
	\draw[-latex] (x11) .. controls +(225:3.5cm) and +(160:2cm) .. (e2);
	\draw[-latex] (e2) .. controls +(95:2cm) and +(295:1cm) .. (x12);
	\draw[-latex] (x32) .. controls +(315:3.5cm) and +(20:2cm) .. (e4);
	\draw[-latex] (e4) .. controls +(85:2cm) and +(245:1cm) .. (x31);
	\draw (a1) +(60:1.5cm) node(a3){$a_3$} +(120:1.5cm) node(a2){$a_2$};
	\foreach \x / \y in {a1/a2,a2/a3,a3/a1,a1/e1,a1/e3}
		{ \draw[-latex] (\x) -- (\y); }
	\draw[-latex] (a1) .. controls +(left:8cm) and +(left:6cm) .. (e2);
	\draw[-latex] (a1) .. controls +(right:8cm) and +(right:6cm) .. (e4);
\end{tikzpicture}
	\caption{Dominance graph from \consref{cons:conp}, example for the formula $(\neg w_1 \vee w_2) \wedge (w_1 \vee \neg w_3)$.}
  \label{fig:up_conp}
\end{figure}
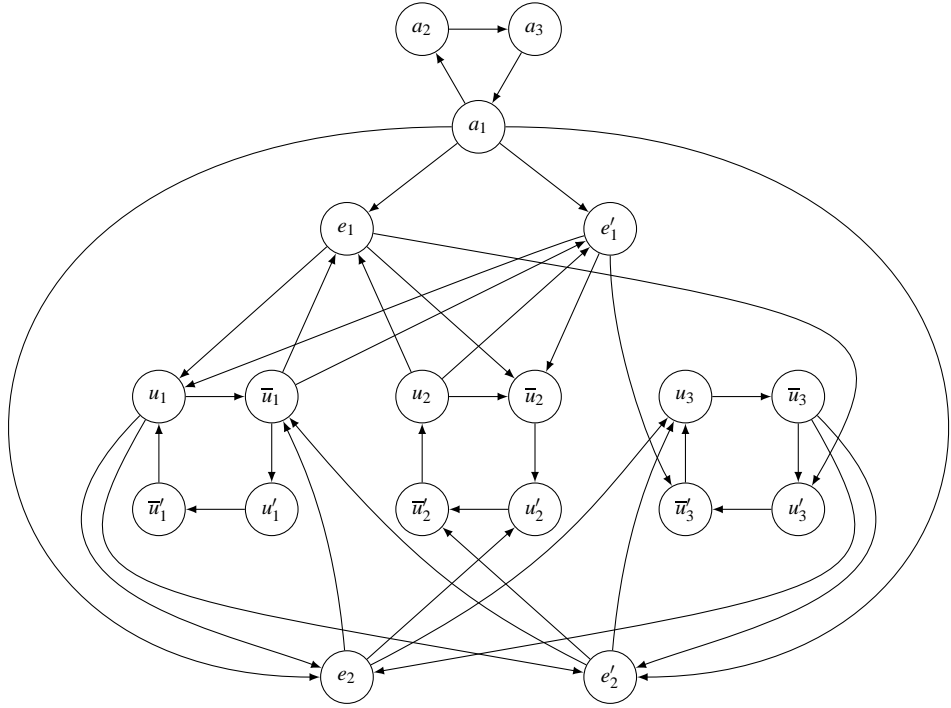

As a more complete example, Figure~\ref{fig:up_conp} shows the
entire dominance graph that corresponds to the concrete formula
$
(\neg w_1 \vee w_2) \wedge (w_1 \vee \neg w_3)$, which can be
satisfied by setting, for example, each of $w_1$, $w_2$, and $w_3$ to
true.  A minimal upward covering set for $A$ corresponding to this assignment
is $M=\{u_1,u'_1,u_2,u'_2,u_3,u'_3,a_1,a_2,a_3\}$.  Note that 
neither $e_1$ nor $e_2$
occurs in~$M$, and none of them occurs in any other minimal upward covering
set for $A$ either.  For alternative $e_1$ this can be seen as follows for the
example shown in Figure~\ref{fig:up_conp}.  If there were a
minimal upward covering set $M'$ for $A$ containing $e_1$ (and thus
also~$e'_1$, since they both are dominated by the same alternatives)
then neither $\overline{u}_1$ nor $u_2$ (which dominate~$e_1$) must
upward cover $e_1$ in~$M'$,
so all alternatives corresponding to the variables
$w_1$ and $w_2$ (i.e., $\{u_i, \overline{u}_i, u_i', \overline{u}_i'
\condition i \in \{1,2\}\}$) would also have to be contained in~$M'$.
Due to $e_1 \succ u'_3$ and $e'_1 \succ \overline{u}'_3$, all
alternatives correponding to $w_3$ (i.e., $\{u_3, \overline{u}_3,
u_3', \overline{u}_3'\}$) are in $M'$ as well. Consequently, $e_2$ and
$e'_2$ are no longer upward covered and must also be in~$M'$.  The
alternatives $a_1,a_2$, and $a_3$ are contained in every minimal
upward covering set
for~$A$.  But then $M'$ is not
minimal because the upward covering set $M$, which corresponds to the
satisfying assignment stated above, is a strict subset of~$M'$.
Hence, $e_1$ cannot be contained in any minimal upward covering set
for~$A$.

\begin{figure}[tb]
	\centering
	\subfigure[$w_i$ occurs in $f_j$ as a positive literal]{
	\label{fig:up_pos} \quad\quad\quad \begin{tikzpicture}[scale=1]
  \tikzstyle{every node}=[circle,draw,minimum size=7mm,inner sep=0pt,font=\scriptsize]
	\draw (0,0) node(x4){$\bu_i'$} ++(0,1.5) node(x1){$u_i$} ++(0,1.5) node(e1){$e_j$} ++(1.5,0) node(e2){$e_j'$} ++(0,-1.5) node(x2){$\bu_i$} ++(0,-1.5) node(x3){$u_i'$};
	\foreach \x / \y in {x1/x2,x2/x3,x3/x4,x4/x1}
		{ \draw[-latex] (\x) -- (\y); }
	\foreach \x / \y in {x1/e1,x1/e2,e1/x2,e2/x2}
		{ \draw[-latex] (\x) -- (\y); }
\end{tikzpicture} \quad\quad\quad} \quad\quad
	\subfigure[$w_i$ occurs in $f_j$ as a negative literal]{
	\label{fig:up_neg} \quad\quad\quad \begin{tikzpicture}[scale=1]
  \tikzstyle{every node}=[circle,draw,minimum size=7mm,inner sep=0pt,font=\scriptsize]
	\draw (0,0) node(x4){$\bu_i'$} ++(0,1.5) node(x1){$u_i$} ++(0,1.5) node(e1){$e_j$} ++(1.5,0) node(e2){$e_j'$} ++(0,-1.5) node(x2){$\bu_i$} ++(0,-1.5) node(x3){$u_i'$};
	\foreach \x / \y in {x1/x2,x2/x3,x3/x4,x4/x1}
		{ \draw[-latex] (\x) -- (\y); }
	\foreach \x / \y in {e1/x1,e2/x1,x2/e1,x2/e2}
		{ \draw[-latex] (\x) -- (\y); }
\end{tikzpicture} \quad\quad\quad} \quad\quad
	\subfigure[$w_i$ does not occur in $f_j$]{
	\label{fig:up_nil} \quad\quad \begin{tikzpicture}[scale=1]
  \tikzstyle{every node}=[circle,draw,minimum size=7mm,inner sep=0pt,font=\scriptsize]
	\draw (0,0) node(x4){$\bu_i'$} ++(0,1.5) node(x1){$u_i$} ++(0,1.5) node(e1){$e_j$} ++(1.5,0) node(e2){$e_j'$} ++(0,-1.5) node(x2){$\bu_i$} ++(0,-1.5) node(x3){$u_i'$};
	\foreach \x / \y in {x1/x2,x2/x3,x3/x4,x4/x1}
		{ \draw[-latex] (\x) -- (\y); }
	\foreach \x / \y in {e1/x3,e2/x4}
		{ \draw[-latex] (\x) -- (\y); }
\end{tikzpicture} \quad\quad}
	\caption{Parts of the dominance graph defined in \consref{cons:conp}.}
	\label{fig:up_gadgets}
\end{figure}
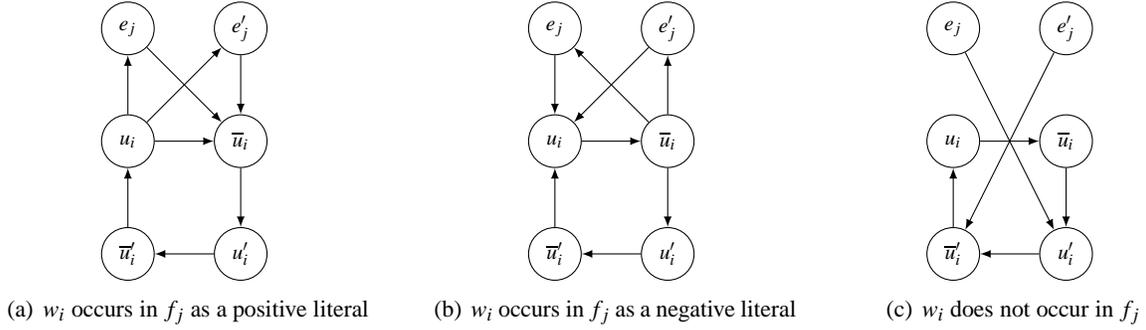

We now show some properties of the dominance graph created by
Construction~\ref{cons:conp} in
general.  We will need these properties for the proofs in
Section~\ref{sec:proofs}.  The first property, stated in
Claim~\ref{cla:key}, has already been seen in the example above.

\begin{claim}
\label{cla:key}
Consider the dominance graph $(A,\succ)$
created by Construction~\ref{cons:conp},
and fix any $j$, $1 \leq j \leq \ell$.  For each minimal upward covering
set $M$ for~$A$, if $M$ contains the alternative $e_j$ then all other
alternatives are contained in $M$ as well (i.e., $A = M$).
\end{claim}

\begin{proofs}To simplify notation, we will prove the
claim only for the case of $j=1$.  However, since there is nothing
special about $e_1$ in our argument, the same property can be shown by
an analogous argument for each~$j$, $1 \leq j \leq \ell$.

Let $M$ be any minimal upward covering set for~$A$, and suppose that
$e_1 \in M$.  First note that the dominators of $e_1$ and $e_1'$ are
always the same (albeit $e_1$ and $e_1'$ may dominate different
alternatives).  Thus, for each minimal upward covering set, either
both $e_1$ and $e_1'$ are contained in it, or they both are not.
Thus, since $e_1 \in M$, we have $e_1' \in M$ as well.

Since the alternatives $a_1$, $a_2$, and $a_3$ form an undominated three-cycle,
they each are contained in every minimal upward covering set for~$A$.
In particular, $\{a_1, a_2, a_3\} \subseteq M$.  Furthermore, no
alternative $e_j$ or $e_j'$, $1 \leq j \leq \ell$, can upward cover any other
alternative in~$M$, because $a_1 \in M$ and $a_1$ dominates $e_j$ and
$e_j'$ but none of the alternatives that are dominated by either $e_j$
or~$e_j'$.  In particular, no alternative in any of the $k$
four-cycles $u_i \succ \overline{u}_i \succ u'_i \succ \overline{u}'_i
\succ u_i$ can be upward covered by any alternative $e_j$ or $e_j'$, and so
they each must be upward covered within their cycle.  For each of these
cycles, every minimal upward covering set for $A$ must contain at least
one of the sets $\{u_i, u'_i\}$ and $\{\overline{u}_i,
\overline{u}'_i\}$, since at least one is needed to upward cover the other
one.\footnote{The argument is analogous to that for the construction
of Brandt and Fischer~\cite{bra-fis:j:minimal-covering-set} in their
proof of Theorem~\ref{thm:up_np}.  However, in contrast with their
construction, which implies that \emph{either $\{x_i, x'_i\}$ or
$\{\overline{x}_i, \overline{x}'_i\}$, $1 \leq i \leq n$, but not
both}, must be contained in any minimal upward covering set for~$A$
(see Figure~\ref{fig:up_np}), our construction also allows for
both $\{u_i, u'_i\}$ and $\{\overline{u}_i, \overline{u}'_i\}$ being
contained in some minimal upward covering set for~$A$.  Informally
stated, the reason is that, unlike the four-cycles in
Figure~\ref{fig:up_np}, our four-cycles $u_i \succ \overline{u}_i
\succ u'_i \succ \overline{u}'_i \succ u_i$ also have incoming edges.}

Since $e_1 \in M$ and by internal stability, we have that no alternative
from $M$ upward covers~$e_1$.  In addition to~$a_1$, the alternatives
dominating $e_1$ are $u_i$ (for each $i$ such that $w_i$ occurs as a
positive literal in~$f_1$) and $\overline{u}_i$ (for each $i$ such
that $w_i$ occurs as a negative literal in~$f_1$).

First assume that, for
some~$i$, $w_i$ occurs as a positive literal in~$f_1$.  Suppose that
$\{u_i, u'_i\} \subseteq M$.  If $\overline{u}'_i \not\in M$ then
$e_1$ would be upward covered by~$u_i$, which is impossible. Thus 
$\overline{u}'_i \in M$.  But then $\overline{u}_i \in M$ as well, since
$u_i$, the only alternative that could upward cover $\overline{u}_i$,
is itself dominated by~$\overline{u}'_i$.  For the latter
argument, recall that $\overline{u}_i$ cannot be upward covered by any
$e_j$ or~$e_j'$.  Thus, we have shown that $\{u_i, u'_i\} \subseteq M$
implies $\{\overline{u}_i, \overline{u}'_i\} \subseteq M$.
Conversely, suppose that $\{\overline{u}_i, \overline{u}'_i\}
\subseteq M$.  Then $u_i'$ is no longer upward covered by $\overline{u}_i$
and hence must be in $M$ as well.  The same holds for the
alternative~$u_i$, so $\{\overline{u}_i, \overline{u}'_i\} \subseteq
M$ implies $\{u_i, u'_i\} \subseteq M$.  Summing up, if $e_1 \in M$
then $\{u_i, u'_i, \overline{u}_i, \overline{u}'_i\} \subseteq M$ for
each $i$ such that $w_i$ occurs as a positive literal in~$f_1$.

By symmetry of the construction, an analogous argument shows that if
$e_1 \in M$ then $\{u_i, u'_i, \overline{u}_i, \overline{u}'_i\}
\subseteq M$ for each $i$ such that $w_i$ occurs as a negative literal
in~$f_1$.

Now, consider any $i$ such that $w_i$ does not occur in~$f_1$.  We
have $e_1 \succ u'_i$ and $e'_1 \succ \overline{u}'_i$.  Again, none
of the sets $\{u_i, u'_i\}$ and $\{\overline{u}_i,\overline{u}'_i\}$
alone can be contained in~$M$, since otherwise either $u_i$ or
$\overline{u}'_i$ would remain upward uncovered.  Thus, $e_1 \in M$ again
implies that $\{u_i, u'_i, \overline{u}_i, \overline{u}'_i\} \subseteq
M$.

Now it is easy to see that, since $\bigcup_{1\leq i \leq k} \{u_i,
u'_i, \overline{u}_i, \overline{u}'_i\} \subseteq M$ and since $a_1$
cannot upward cover any of the $e_j$ and $e_j'$, $1 \leq j \leq \ell$,
external stability of $M$ enforces that $\bigcup_{1 < j \leq \ell}
\{e_j, e_j'\} \subseteq M$.  Summing up, we have shown that if $e_1$
is contained in any minimal upward covering set~$M$ for~$A$, then
$M=A$.~\end{proofs}
\begin{claim}
\label{cla:no-ej}
Consider Construction~\ref{cons:conp}.
The boolean formula $\varphi$ is satisfiable if and only if there is
no minimal upward covering set for $A$ that contains any of the~$e_j$,
$1\leq j \leq \ell$.
\end{claim}
\begin{proofs} It is enough to prove the claim for the
case $j=1$, since the proof for the other cases is analogous.

From left to right, suppose there is a satisfying assignment $\alpha :
W \rightarrow \{0,1\}$ for~$\varphi$.  Define the set
\[
B_{\alpha} = \{a_1, a_2, a_3\} \cup \{u_i, u'_i \condition \alpha(w_i) = 1\}
\cup \{\overline{u}_i, \overline{u}'_i \condition \alpha(w_i) = 0\}.
\]

Since every upward covering set for $A$ must contain $\{a_1, a_2,
a_3\}$ and at least one of the sets $\{u_i, u'_i\}$ and
$\{\overline{u}_i, \overline{u}'_i\}$ for each~$i$, $1 \leq i \leq k$,
$B_{\alpha}$ is a (minimal) upward covering set for~$A$.  Let $M$ be an
arbitrary minimal upward covering set for~$A$.  By
Claim~\ref{cla:key}, if $e_1$ were contained in~$M$, we would have
$M = A$.  But since $B_{\alpha} \subset A = M$, this contradicts the minimality
of~$M$.  Thus $e_1 \not\in M$. 
(Note that every minimal upward covering set for $A$ obtained from any
satisfying assignment for $\varphi$ contains
exactly $2k+3$ alternatives, and there is no minimal upward
covering set of smaller size for $A$ when $\varphi$ is
satisfiable. This observation will be used later on.)

From right to left, let $M$ be an arbitrary minimal upward covering
set for~$A$ and suppose $e_1 \not\in M$.  By Claim~\ref{cla:key}, if
any of the $e_j$, $1 < j \leq \ell$, were contained in~$M$, it would
follow that $e_1 \in M$, a contradiction.  Thus, $\{e_j \condition 1
\leq j \leq \ell\} \cap M = \emptyset$. It follows that each $e_j$
must be upward covered by some alternative in~$M$.  It is easy to see that
for each $j$, $1 \leq j \leq \ell$, and for each $i$, $1 \leq i \leq
k$, $e_j$ is upward covered in $M\cup\{e_j\}\supseteq\{u_i,u'_i\}$ if $w_i$
occurs in $f_j$ as a positive literal, and $e_j$ is upward covered in 
$M\cup\{e_j\}\supseteq \{\overline{u}_i, \overline{u}'_i\}$ if $w_i$ occurs in
$e_j$ as a negative literal.  It can never be the case that all four
alternatives, $\{u_i, u'_i, \overline{u}_i, \overline{u}'_i\}$, are
contained in~$M$, because then either $e_j$ would no longer be upward covered
in $M$ or the resulting set $M$ was not minimal.  Now, $M$ induces a
satisfying assignment for $\varphi$ by setting, for each $i$, $1 \leq
i \leq k$, $\alpha(w_i) = 1$ if $u_i \in M$, and $\alpha(w_i) = 0$ if
$\overline{u}_i \in M$.~\end{proofs}

\begin{claim}
\label{cla:notsatisfiable-all}
Consider Construction~\ref{cons:conp}.
The boolean formula $\varphi$ is not satisfiable if and only if there
is a unique minimal upward
covering set for~$A$.
\end{claim}

\begin{proofs}
Without loss of generality, we may assume that if $\varphi$ is
satisfiable then it has at least two satisfying assignments.  This can
be ensured, if needed, by adding dummy variables.

From left to right, suppose there is no satisfying assignment for~$\varphi$.
By Claim~\ref{cla:no-ej}, there must be a minimal
upward covering set for $A$
containing one of the $e_j$, $1 \leq j \leq \ell$,
and by Claim~\ref{cla:key} this minimal upward covering set for $A$ must
contain all alternatives. By reason of minimality, there cannot be
another minimal upward covering set for~$A$.

From right to left, suppose there is a unique minimal upward covering
set for $A$. Due to our assumption that if $\varphi$ is satisfiable
then there are at least two satisfying assignments, $\varphi$ cannot be
satisfiable, since if it were, there would be two distinct minimal upward
covering sets corresponding to these assignments (as argued in
the proof of Claim~\ref{cla:no-ej}).~\end{proofs}
Wagner provided a sufficient condition for proving
$\Theta_2^p$-hardness that was useful in
various other contexts (see, e.g.,
\cite{wag:j:min-max,hem-hem-rot:j:dodgson,hem-rot:j:max-independent-set-by-greed,hem-wec:j:mee,hem-rot-spa:j:vcgreedy})
and is stated here as Lemma~\ref{lem:wagner-theta-2-p}.

\begin{lemma}[Wagner \cite{wag:j:min-max}]
\label{lem:wagner-theta-2-p}
Let $S$ be some $\np$-complete problem, and let $T$ be any set. If
there exists a polynomial-time computable function $f$ such that, for
all $m \geq 1$ and all strings $x_1, x_2, \dots,x_{2m}$ satisfying that
if $x_j \in S$ then $x_{j-1} \in S$, $1 < j \leq 2m$, we have
\begin{eqnarray}
\label{eqn:wagner}
\|\{i\condition x_i \in S\}\| \text{ is odd } 
& \Longleftrightarrow & 
f(x_1,x_2,\dots,x_{2m}) \in T,
\end{eqnarray}
then $T$ is $\Theta_2^p$-hard.
\end{lemma}

We will apply Lemma~\ref{lem:wagner-theta-2-p} as well.  In contrast
with those previous results, however, one subtlety in our construction
is due to the fact that we consider not only minimum-size but also
\mbox{(inclusion-)}\allowbreak{minimal} covering sets.  To the best of our 
knowledge, our constructions for the first time apply Wagner's
technique~\cite{wag:j:min-max} to problems defined in terms of 
minimality/maximality rather than minimum/maximum size of a 
solution:\footnote{For example,
recall Wagner's $\Theta_2^p$-completeness result
for testing whether the size of a maximum clique in a given graph is
an odd number~\cite{wag:j:min-max}.  One key ingredient in his proof
is to define an associative operation on graphs, $\bowtie$, such that
for any two graphs $G$ and~$H$, the size of a maximum clique in $G
\bowtie H$ equals the sum of the sizes of a maximum clique in $G$ and
one in~$H$.  This operation is quite simple: Just connect every vertex
of $G$ with every vertex of~$H$.  In contrast, since minimality for
minimal upward covering sets is defined in terms of set inclusion, it
is not at all obvious how to define a similarly simple operation on
dominance graphs such that the minimal upward covering sets in the
given graphs are related to the minimal upward covering sets in the
connected graph in a similarly useful way.}
In Construction~\ref{cons:theta} below, we define a dominance
graph based on Construction~\ref{cons:conp} and the construction
presented in the proof sketch of
Theorem~\ref{thm:up_np} such that
Lemma~\ref{lem:wagner-theta-2-p} can be applied to prove $\mucmember$
$\Theta_2^p$-hard (see Theorem~\ref{thm:minimal-in-some}),
making use of the properties established in
Claims~\ref{cla:key}, \ref{cla:no-ej},
and~\ref{cla:notsatisfiable-all}.

\begin{construction}[For applying Lemma~\ref{lem:wagner-theta-2-p} 
to upward covering set problems]~
\label{cons:theta}
We apply Wagner's Lemma with the $\np$-complete problem $S = \sat$ and
construct a dominance graph.  Fix an arbitrary $m \geq 1$ and let
$\varphi_1, \varphi_2, \dots, \varphi_{2m}$ be $2m$ boolean formulas
in conjunctive normal form
such that if $\varphi_j$ is satisfiable then so is $\varphi_{j-1}$,
for each~$j$, $1 < j \leq 2m$.  Without loss of generality, we assume
that for each~$j$, $1 \leq j \leq 2m$, the first variable of
$\varphi_{j}$ does not occur in all clauses of $\varphi_{j}$.
Furthermore we require $\varphi_j$ to have at least two unsatisfied
clauses if $\varphi_j$ is not satisfiable, and to have at least two
satisfying assignments if $\varphi_j$ is satisfiable.  It is easy to
see that if $\varphi_{j}$ does not have these properties, it can be
transformed into a formula that does have them, without affecting the
satisfiability of the formula.

We will now define a polynomial-time computable function~$f$, which
maps the given $2m$ boolean formulas to 
a dominance graph $(A,\succ)$ with useful properties for upward
covering sets.
Define $A =
\bigcup_{j=1}^{2m} A_j$ and the dominance relation $\succ$ on $A$ by
\[
\left(\bigcup_{j=1}^{2m} \succ_j
\right) \cup \left(\bigcup_{i=1}^{m} \left\{(u_{1,2i}',d_{2i-1}),
(\overline{u}_{1,2i}',d_{2i-1})\right\} \right) \cup
\left(\bigcup_{i=2}^{m} \left\{(d_{2i-1}, z) \condition z \in
A_{2i-2}\right\} \right),
\]
where we use the following notation:
\begin{enumerate}
\item For each~$i$, $1 \leq i \leq m$, let $(A_{2i-1},\succ_{2i-1})$
be the dominance graph that results from the formula $\varphi_{2i-1}$
according to Brandt and Fischer's construction given in the proof
sketch of Theorem~\ref{thm:up_np}.  We use the same names for the
alternatives in $A_{2i-1}$
as in that proof sketch, except that we attach the
subscript $2i-1$.  For example, alternative $d$ from the proof sketch
of Theorem~\ref{thm:up_np} now becomes $d_{2i-1}$, $x_1$ becomes
$x_{1,2i-1}$, $y_1$ becomes $y_{1,2i-1}$, and so on.

\item For each~$i$, $1 \leq i \leq m$, let $(A_{2i},\succ_{2i})$ be
the dominance graph that results from the formula $\varphi_{2i}$
according to Construction~\ref{cons:conp}.
We use the same names for the alternatives in $A_{2i}$
as in that construction, except that we attach the subscript $2i$.  For example,
alternative $a_1$ from Construction~\ref{cons:conp}
now becomes
$a_{1,2i}$, $e_1$ becomes $e_{1,2i}$, $u_1$ becomes $u_{1,2i}$, and so
on.

\item \label{cons:theta:item-three}
For each $i$, $1 \leq i \leq m$, connect the dominance graphs
$(A_{2i-1},\succ_{2i-1})$ and $(A_{2i},\succ_{2i})$ as follows.  Let
$u_{1,2i}, \overline{u}_{1,2i}, u_{1,2i}', \overline{u}_{1,2i}' \in
A_{2i}$ be the four alternatives in the cycle corresponding to the
first variable of $\varphi_{2i}$.  Then both $u_{1,2i}'$ and
$\overline{u}_{1,2i}'$ dominate $d_{2i-1}$.  The resulting dominance
graph is denoted by $(B_i, \succ_i^B)$.

\item Connect the $m$ dominance graphs $(B_i, \succ_i^B)$, $1 \leq i
\leq m$, as follows: For each $i$, $2 \leq i \leq m$, $d_{2i-1}$
dominates all alternatives in $A_{2i-2}$.
\end{enumerate}
\end{construction}

\begin{figure}[tb]
  \centering
	\begin{tikzpicture}[scale=.8]
	\scriptsize
	\tikzstyle{every circle node}=[draw,minimum size=7mm,inner sep=0pt,font=\scriptsize]
	\draw (0,0) node(d1)[circle]{$d_1$} +(-1,1) rectangle +(1,-4) node[above left]{$A_1$} ++(3,0) +(-1,1) rectangle +(1,-4) node[above left]{$A_2$} +(1,0) coordinate(A2) +(0,-1) node(u12)[circle]{$u_{1,2}'$} +(0,-2.5) node(bu12)[circle]{$\bu_{1,2}'$} ++(3,0) node(d2)[circle]{$d_3$} +(-1,1) rectangle +(1,-4) node[above left]{$A_3$} ++(2.5,0) +(-1,1) rectangle +(1,-4) node[above left]{$A_4$} +(1,0) coordinate(A4) +(0,-1) node(u14)[circle]{$u_{1,4}'$} +(0,-2.5) node(bu14)[circle]{$\bu_{1,4}'$} ++(2,0) node(d){$\dots$} ++(2,0) node(d5)[circle]{$d_{2m\text{-}1}$} +(-1,1) rectangle +(1,-4) node[above left]{$A_{2m-1}$} ++(2.5,0) +(-1,1) rectangle +(1,-4) node[above left]{$A_{2m}$} +(0,-1) node(u16)[circle]{$u_{1,2m}'$} +(0,-2.5) node(bu16)[circle]{$\bu_{1,2m}'$};
	\foreach \x / \y in {u12/d1,bu12/d1,d2/A2,u14/d2,bu14/d2,d/A4,u16/d5,bu16/d5} {
		\draw[-latex] (\x) -- (\y); }
	\draw	(d5) --(d);
\end{tikzpicture}
	\caption{Dominance graph from Construction~\ref{cons:theta}.  Most alternatives, and all
	edges between pairs of alternatives, in $A_j$ for $1 \leq j \leq 2m$ have been omitted. All edges between alternatives in $A_i$ and alternatives in $A_j$ for $i \neq j$ are shown.  An edge incident to a set of alternatives represents an edge incident to \emph{each} alternative in the set.}
  \label{fig:up_ttp}
\end{figure}
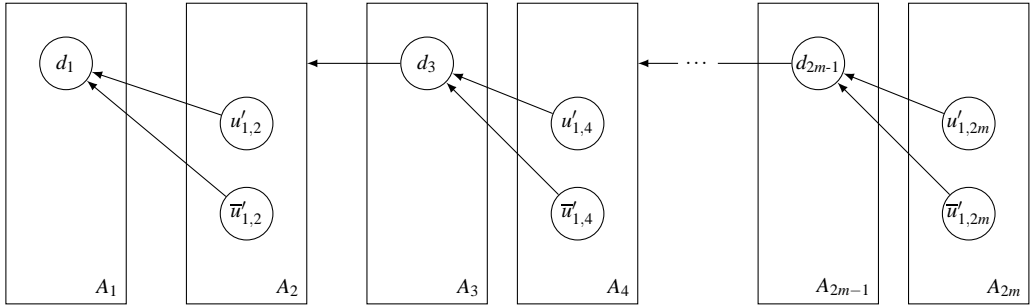

The dominance graph $(A,\succ)$ is sketched in Figure~\ref{fig:up_ttp}.
Clearly,
$(A,\succ)$
is computable in polynomial time.

Before we use this construction to obtain $\Theta_2^p$-hardness results
for some of our upward covering set problems in Section~\ref{sec:proofs},
we will again show some useful properties of the constructed dominance
graph, and we 
first consider the dominance graph $(B_i, \succ_i^B)$ (see
Step~\ref{cons:theta:item-three} in Construction~\ref{cons:theta})
separately,\footnote{Note that our argument about $(B_i, \succ_i^B)$
  can be used to show, in effect, $\DP$-hardness of
  upward covering set problems, where $\DP$ is the class of
  differences of any two $\np$ sets~\cite{pap-yan:j:dp}.  Note that
  $\DP$ is the second level of the boolean hierarchy over $\np$ (see
  Cai et
  al.~\cite{cai-gun-har-hem-sew-wag-wec:j:bh1,cai-gun-har-hem-sew-wag-wec:j:bh2}),
  and it holds that $\np \cup \conp \subseteq \DP \subseteq
  \Theta_2^p$.  Wagner~\cite{wag:j:min-max} proved appropriate analogs
  of Lemma~\ref{lem:wagner-theta-2-p} for each level of the boolean
  hierarchy.  In particular, the analogous criterion for
  $\DP$-hardness is obtained by using the wording of
  Lemma~\ref{lem:wagner-theta-2-p} except with the value of $m=1$
  being fixed.}  for any fixed $i$ with $1 \leq i \leq m$.  Doing so
will simplify our argument for the whole dominance graph $(A,\succ)$.
Recall that $(B_i,\succ_i^B)$ results from the formulas
$\varphi_{2i-1}$ and $\varphi_{2i}$.
\begin{claim}
\label{cla:bi}
Consider Construction~\ref{cons:theta}.
Alternative $d_{2i-1}$ is contained in some minimal upward covering set for
$(B_i,\succ_i^B)$ if and only if $\varphi_{2i-1}$ is satisfiable and
$\varphi_{2i}$ is not satisfiable.
\end{claim}

\begin{proofs}
Distinguish the following three cases.

\begin{casedistinction}
\item \label{case:one} $\varphi_{2i-1} \in \sat$ and $\varphi_{2i} \in
\sat$.  Since $\varphi_{2i}$ is satisfiable, it follows from the proof
of 
Claim~\ref{cla:key} that for each minimal upward covering set $M$
for
$(B_i,\succ_i^B)$,
either $\{u_{1,2i}, u'_{1,2i}\} \subseteq M$ or
$\{\overline{u}_{1,2i}, \overline{u}'_{1,2i}\} \subseteq M$, but not
both, and that none of the $e_{j,2i}$ and $e'_{j,2i}$ is in~$M$.  If
$\overline{u}'_{1,2i} \in M$ but $u'_{1,2i} \not\in M$, then $d_{2i-1}
\not\in \mathrm{UC}_u(M)$, since $\overline{u}'_{1,2i}$ upward covers
$d_{2i-1}$ within~$M$.  If $u'_{1,2i} \in M$ but $\overline{u}_{1,2i}
\not\in M$, then $d_{2i-1} \not\in \mathrm{UC}_u(M)$, since
$u'_{1,2i}$ upward covers $d_{2i-1}$ within~$M$.  Hence, by
internal stability, $d_{2i-1}$ is not contained in~$M$.

\item \label{case:two} $\varphi_{2i-1} \not\in \sat$ and $\varphi_{2i}
\not\in \sat$.  Since $\varphi_{2i-1} \not\in \sat$, it follows from
the proof of Theorem~\ref{thm:up_np} that each minimal upward covering
set $M$ for $(B_i,\succ_i^B)$ contains at least one alternative
$y_{j,2i-1}$ (corresponding to some clause of $\varphi_{2i-1}$) that
upward covers $d_{2i-1}$.  Thus $d_{2i-1}$ cannot be in~$M$, again by
internal stability.

\item \label{case:three} $\varphi_{2i-1} \in \sat$ and $\varphi_{2i}
\not\in \sat$.  Since $\varphi_{2i-1} \in \sat$, it follows from the
proof of Theorem~\ref{thm:up_np} that there exists a minimal upward
covering set $M'$ for $(A_{2i-1}, \succ_{2i-1})$ that corresponds to
a satisfying truth assignment for $\varphi_{2i-1}$.  In particular,
none of the $y_{j,2i-1}$ is in~$M'$.  On the other hand, since
$\varphi_{2i} \not\in \sat$, it follows from 
Claim~\ref{cla:notsatisfiable-all}
that $A_{2i}$ is the only minimal upward
covering set for $(A_{2i}, \succ_{2i})$.  Define $M = M' \cup A_{2i}$.
It is easy to see that $M$ is a minimal upward covering set for
$(B_i,\succ_i^B)$, since
the only edges between $A_{2i-1}$ and $A_{2i}$ are those from
$\overline{u}'_{1,2i}$ and $u'_{1,2i}$ to $d_{2i-1}$,
and both $\overline{u}'_{1,2i}$ and $u'_{1,2i}$ are dominated by elements
in $M$ not dominating~$d_{2i-1}$.

We now show that $d_{2i-1} \in M$.  Note that $\overline{u}'_{1,2i}$,
$u'_{1,2i}$, and the $y_{j,2i-1}$ are the only alternatives in $B_i$
that dominate $d_{2i-1}$.  Since none of the $y_{j,2i-1}$ is in $M$,
they do not upward cover $d_{2i-1}$.  Also, $u'_{1,2i}$ doesn't upward
cover $d_{2i-1}$, since $\overline{u}_{1,2i} \in M$ and
$\overline{u}_{1,2i}$ dominates $u'_{1,2i}$ but not $d_{2i-1}$.  On
the other hand, by our assumption that the first variable of
$\varphi_{2i}$ does not occur in all clauses, there exist alternatives
$e_{j,2i}$ and $e'_{j,2i}$ in $M$ that dominate $\overline{u}'_{1,2i}$
but not $d_{2i-1}$, so $\overline{u}'_{1,2i}$ doesn't upward cover
$d_{2i-1}$ either.  Thus $d_{2i-1} \in M$.
\end{casedistinction}
Note that, by our assumption on how the formulas are ordered, the
fourth case (i.e., $\varphi_{2i-1} \not\in \sat$ and $\varphi_{2i} \in
\sat$) cannot occur.  Thus, the proof is complete.~\end{proofs}

\begin{claim}
\label{cla:eachM_i-is-included}
Consider Construction~\ref{cons:theta}.
For each~$i$, $1 \leq i \leq m$, let $M_i$ be the minimal upward
covering set for $(B_i,\succ_i^B)$ according to the cases in the
proof of Claim~\ref{cla:bi}.  Then
each of the sets $M_i$ must be contained in every minimal upward
covering set for $(A,\succ)$.
\end{claim}

\begin{proofs}
The minimal upward covering set $M_m$ for $(B_m,\succ_m^B)$ must
be contained in every minimal upward covering set for $(A,\succ)$,
since no alternative in $A - B_m$ dominates any alternative in~$B_m$.
On the other hand, for each~$i$, $1 \leq i < m$, no alternative in
$B_i$ can be upward covered by $d_{2i+1}$ (which is the only element
in $A - B_i$ that dominates any of the elements of~$B_i$), since
$d_{2i+1}$ is dominated within every minimal upward covering set for
$B_{i+1}$ (and, in particular, within $M_{i+1}$).  Thus, each of the
sets $M_i$, $1 \leq i \leq m$, must be contained in every minimal
upward covering set for $(A,\succ)$.~\end{proofs}

\begin{claim}
\label{cla:wagner-mucm}
Consider Construction~\ref{cons:theta}.  It holds that
\begin{eqnarray}
\label{eqn:wagner-mucm}
\|\{i\condition \varphi_i \in \sat\}\| \text{ is odd} 
& \Longleftrightarrow & 
\mbox{$d_1$ is contained in some minimal upward covering set $M$ for $A$}.
\end{eqnarray}
\end{claim}

\begin{proofs}
To show (\ref{eqn:wagner-mucm}) from left to right, suppose
$\|\{i\condition \varphi_i \in \sat\}\|$ is odd.  Recall that for
each~$j$, $1 < j \leq 2m$, if $\varphi_j$ is satisfiable then so is
$\varphi_{j-1}$.  Thus, there exists some~$i$, $1 \leq i \leq m$, such
that $\varphi_1, \dots , \varphi_{2i-1} \in \sat$ and $\varphi_{2i},
\dots , \varphi_{2m} \not\in \sat$.  In Case~\ref{case:three}
in the proof of Claim~\ref{cla:bi}
we have seen that there is some minimal upward
covering set for $(B_i,\succ_i^B)$---call it~$M_i$---that corresponds
to a satisfying assignment of $\varphi_{2i-1}$ and that contains all
alternatives of $A_{2i}$.  In particular, $M_i$ contains $d_{2i-1}$.
For each $j \neq i$, $1 \leq j \leq m$, let $M_j$ be some minimal
upward covering set for $(B_j,\succ_j^B)$ according to
Case~\ref{case:one} (if $j<i$) and Case~\ref{case:two} (if $j>i$)
in the proof of Claim~\ref{cla:bi}.

In Case~\ref{case:one} we have seen that $d_{2i-3}$ is upward covered
either by $\overline{u}'_{1,2i-3}$ or by $u'_{1,2i-3}$.  This is no
longer the case, since $d_{2i-1}$ is in $M_i$ and it dominates all
alternatives in $A_{2i-2}$ but not $d_{2i-3}$. By assumption,
$\varphi_{2i-3}$ is satisfiable, so there exists a minimal upward
covering set, which contains $d_{2i-3}$ as well.  Thus, setting 
\[
M = \{d_1, d_3, \dots , d_{2i-1}\} \cup \bigcup_{1 \leq j \leq m} M_j ,
\]
it follows that $M$ is a minimal upward covering set for
$(A,\succ)$ containing~$d_1$.

To show (\ref{eqn:wagner-mucm}) from right to left, suppose that
$\|\{i\condition \varphi_i \in \sat\}\|$ is even.  For a
contradiction, suppose that there exists some minimal upward covering
set $M$ for $(A,\succ)$ that contains~$d_1$.  If $\varphi_1 \not\in
\sat$ then we immediately obtain a contradiction by the argument in
the proof of Theorem~\ref{thm:up_np}.  On the other hand, if $\varphi_1
\in \sat$ then our assumption that $\|\{i\condition \varphi_i \in
\sat\}\|$ is even implies that $\varphi_2 \in \sat$.  It follows from
the proof of 
Claim~\ref{cla:key}
that every minimal upward covering
set for $(A,\succ)$ (thus, in particular, $M$) contains either
$\{u_{1,2i}, u'_{1,2i}\}$ or $\{\overline{u}_{1,2i},
\overline{u}'_{1,2i}\}$, but not both, and that none of the $e_{j,2i}$
and $e'_{j,2i}$ is in~$M$.  By the argument presented in Case~3
in the proof of Claim~\ref{cla:bi},
the only way to prevent $d_1$ from being upward covered by an element of~$M$,
either $u_{1,2}'$ or $\overline{u}_{1,2}'$, is to include $d_3$ 
in $M$ as well.\footnote{This implies that $d_1$ is not
upward covered by either $u_{1,2}'$ or $\overline{u}_{1,2}'$, since $d_3$
dominates them both but not~$d_1$.}  By applying the same argument
$m-1$ times, we
will eventually reach a contradiction, since $d_{2m-1} \in M$ can no
longer be prevented from being upward covered by an element of~$M$, either
$u_{1,2m}'$ or $\overline{u}_{1,2m}'$.  Thus, no minimal upward
covering set $M$ for $(A,\succ)$ contains~$d_1$, which
completes the proof of~(\ref{eqn:wagner-mucm}).~\end{proofs}

Furthermore, it holds that $\|\{i\condition \varphi_i \in \sat\}\|$ is
odd if and only if $d_1$ is contained in all minimum-size upward
covering sets for~$A$.
This is true since the minimal upward covering sets for $A$ that contain
$d_1$ are those that correspond to some satisfying assignment for all
satisfiable formulas~$\varphi_i$, and as we have
seen in the analysis of Construction~\ref{cons:conp} and the proof
sketch of Theorem~\ref{thm:up_np}, these are the minimum-size upward 
covering sets for~$A$.

\subsection{Minimal and Minimum-Size Downward Covering Sets}

Turning now to the constructions used to show complexity results
about minimal/minimum-size downward covering sets,
we will again start by giving a proof sketch of a result
due to 
Brandt and Fischer~\cite{bra-fis:j:minimal-covering-set}, since the
following constructions and proofs are based on their construction
and proof.

\begin{theorem}[Brandt and Fischer~\cite{bra-fis:j:minimal-covering-set}] 
\label{thm:down_np}
Deciding whether a designated alternative is contained in some minimal
downward covering set for a given dominance graph 
is $\np$-hard (i.e., $\mdcmember$ is  $\np$-hard),
even if a downward covering set is guaranteed to exist.
\end{theorem}

\begin{proofsketch}
  $\np$-hardness of $\mdcmember$
  is again shown by a reduction from $\sat$. Given
  a boolean formula  in conjunctive normal form,
  $\varphi(v_1, v_2, \dots , v_n) = c_1 \wedge c_2
  \wedge \dots \wedge c_r$, over the set $V = \{v_1, v_2, \dots ,
  v_n\}$ of variables, construct 
a dominance graph $(A,\succ)$
  as follows. The set of alternatives is
\[
A= \{x_i,\overline{x}_i, x_i',\overline{x}_i',x_1'',\overline{x}_i''
\condition v_i \in V\}
\cup
\{y_j,z_j \condition c_j \mbox{ is a clause in } \varphi \}
\cup
\{d\},
\]
where the membership of alternative $d$ in a minimal downward covering
set is to be decided.
The dominance relation $\succ$ is defined as follows:
\begin{itemize}
\item For each $i$, $1 \leq i \leq n$,  there is a cycle $x_i
  \succ \overline{x}_i \succ x_i' \succ \overline{x}_i' \succ x_i''
  \succ \overline{x}_i'' \succ x_i$ with two nested three-cycles,
  $x_i \succ x_i' \succ x_i'' \succ x_i$ and $\overline{x}_i \succ
  \overline{x}_i' \succ \overline{x}_i'' \succ \overline{x}_i$;
\item if variable $v_i$ occurs in clause $c_j$ as a
  positive literal, then $y_j \succ x_i$;
\item if variable $v_i$ occurs in clause
  $c_j$ as a negative literal, then $y_j \succ \overline{x}_i$;
\item for each $j$, $1 \leq j \leq r$, we have
  $d \succ y_j$ and $z_j \succ d$; and
\item for each $i$ and $j$ with $1 \leq i,j \leq r$ and $i \neq j$,
we have $z_i \succ y_j$.
\end{itemize}

Brandt and Fischer~\cite{bra-fis:j:minimal-covering-set} showed that
there is a minimal downward covering set containing $d$ if and only if
$\varphi$ is satisfiable. An example of this reduction is shown in
Figure~\ref{fig:down_np} for the boolean formula $(v_1 \vee \neg v_2 \vee v_3) \wedge (\neg v_1 \vee \neg v_3)$.  The set $\{x_1,x_1',x_1'',x_2,x_2',x_2'',\bx_3,\bx_3',\bx_3'',y_1,y_2,z_1,z_2,d\}$
is a minimal downward covering set for the dominance graph shown in
Figure~\ref{fig:down_np}.  This
set corresponds to the truth assignment that sets $v_1$ and $v_2$ to true and $v_3$ to false, and it contains the designated
alternative~$d$.~\end{proofsketch}

\begin{figure}[tb]
  \centering
	\begin{tikzpicture}[scale=.8]
  \tikzstyle{every node}=[circle,draw,minimum size=7mm,inner sep=0pt,font=\scriptsize]
	\foreach \i in {1,2,3} {
		\draw (5*\i,0) node(x\i1){$x_{\i}$} ++(0:1.5cm) node(y\i1){$\bx_{\i}$} ++(300:1.5cm) node(x\i2){$x_{\i}'$} ++(240:1.5cm) node(y\i2){$\bx_{\i}'$} ++(180:1.5cm) node(x\i3){$x_{\i}''$} ++(120:1.5cm) node(y\i3){$\bx_{\i}''$};
		\foreach \x / \y in {x\i1/y\i1,y\i1/x\i2,x\i2/y\i2,y\i2/x\i3,x\i3/y\i3,y\i3/x\i1,x\i1/x\i2,x\i2/x\i3,x\i3/x\i1,y\i1/y\i2,y\i2/y\i3,y\i3/y\i1}
			{ \draw[-latex] (\x) -- (\y); }}
	\path (y11) -- node[draw,above=1.5cm](y1){$y_1$} (x21);
	\path (y21) -- node[draw,above=1.5cm](y2){$y_2$} (x31);
	\draw (y1) +(0,1.5) node(z2){$z_2$} (y2) +(0,1.5) node(z1){$z_1$};
	\path (z2) -- node[draw](d){$d$} (z1);
	\foreach \x / \y in {y1/x11,y1/y21,y1/x31,y2/y11,y2/y31,z2/y1,z2/d,z1/y2,z1/d,d/y1,d/y2}
		{ \draw[-latex] (\x) -- (\y); }
\end{tikzpicture}
	\caption{Dominance graph for \thmref{thm:down_np}, example for the formula $(v_1 \vee \neg v_2 \vee v_3) \wedge (\neg v_1 \vee \neg v_3)$.}
  \label{fig:down_np}
\end{figure}

Regarding their construction sketched above, Brandt and
Fischer~\cite{bra-fis:j:minimal-covering-set} showed that every
minimal downward covering set for $A$ must contain exactly three
alternatives for every variable $v_i$ (either $x_i$, $x_i'$,
and~$x_i''$, or $\overline{x}_i$, $\overline{x}_i'$,
and~$\overline{x}_i''$), and the undominated alternatives
$z_1, \ldots, z_r$.  Thus, each minimal downward covering set for $A$
consists of at least $3n+r$ alternatives and induces a truth assignment
$\alpha$ for~$\varphi$. 
The number of alternatives contained in any minimal
downward covering set for $A$ corresponding to an assignment $\alpha$ is
$3n+r+k$, where $k$ is the number of clauses that are satisfied if
$\alpha$ is an assignment not satisfying~$\varphi$, and where $k=r+1$
if $\alpha$ is a satisfying assignment for~$\varphi$.
As a consequence, minimum-size downward covering
sets for $A$ correspond to those assignments for $\varphi$
that satisfy the least
possible number of clauses of~$\varphi$.\footnote{This is different from
the case of minimum-size \emph{upward} covering sets for the
dominance graph constructed in the proof sketch of
Theorem~\ref{thm:up_np}.  The construction in the proof sketch of
Theorem~\ref{thm:down_np} cannot be used to obtain
complexity results for minimum-size downward covering sets in the same
way as the construction in 
the proof sketch of Theorem~\ref{thm:up_np} was used to obtain
complexity results for minimum-size upward covering sets.}

Next, we provide a different construction to transform a
given boolean formula into a dominance graph.
This construction will later be merged with the construction
from the proof sketch of Theorem~\ref{thm:down_np} so as to apply
Lemma~\ref{lem:wagner-theta-2-p} to downward covering set problems.

\begin{construction}[To be used for showing NP- and coNP-hardness for 
downward covering set problems]~
\label{cons:down-conp}
Given a boolean formula in conjunctive normal form,
$\varphi(w_1,w_2,\dots,w_k)=f_1 \wedge f_2 \wedge
\dots \wedge f_\ell$, over the set $W=\{w_1,
w_2, \dots , w_k\}$ of variables, we construct a dominance graph
$(A,\succ)$. The set of alternatives is
$$ A= A_1 \cup A_2 \cup \{\ha \condition a \in A_1 \cup A_2\} \cup
\{b,c,d\} $$
with
$A_1=\{x_i,x_i',x_i'',\overline{x}_i,\overline{x}_i',\overline{x}_i'',z_i,z_i',z_i''
\condition w_i \in W\}$ and $A_2=\{y_j \condition f_j \text{ is a
clause in } \varphi\}$, and the dominance relation $\succ$is defined by:
\begin{itemize}
\item For each $i$, $1 \leq i \leq k$, there is, similarly to the
  construction in the proof of Theorem~\ref{thm:down_np}, a cycle $x_i
  \succ \overline{x}_i \succ x_i' \succ \overline{x}_i' \succ x_i''
  \succ \overline{x}_i'' \succ x_i$ with two nested three-cycles,
  $x_i \succ x_i' \succ x_i'' \succ x_i$ and $\overline{x}_i \succ
  \overline{x}_i' \succ \overline{x}_i'' \succ \overline{x}_i$, and
  additionally we have $z_i' \succ z_i \succ x_i$, $z_i'' \succ z_i \succ
  \overline{x_i}$, $z_i' \succ x_i$, $z_i'' \succ \overline{x}_i$, and
  $ d\succ z_i$;
\item if variable $w_i$ occurs in clause $f_j$ as a positive literal, then $x_i \succ
  y_j$;
\item if variable $w_i$ occurs in clause $f_j$ as a negative literal, then
  $\overline{x}_i \succ y_j$;
\item for each $a \in A_1 \cup A_2$, we have $b \succ \ha$, $a \succ
  \ha $, and $\ha \succ d$;
\item for each $j$, $1 \leq j \leq \ell$, we have $d \succ y_j$; and
\item $c \succ d$.
\end{itemize}
\end{construction}

\begin{figure}
\centering
\begin{tikzpicture}[scale=.8]
  \scriptsize
	\tikzstyle{every circle node}=[draw,minimum size=7mm,inner sep=0pt,font=\scriptsize]
	\foreach \i in {1,2,3} {
		\draw (5*\i,0) node(x\i1)[circle]{$x_{\i}$} ++(60:1.5cm) node(z\i3)[circle]{$z_{\i}$} +(180:1.5cm) node(z\i1)[circle]{$z_{\i}'$} +(0:1.5cm) node(z\i2)[circle]{$z_{\i}''$} ++(300:1.5cm) node(y\i1)[circle]{$\bx_{\i}$} ++(300:1.5cm) node(x\i2)[circle]{$x_{\i}'$} ++(240:1.5cm) node(y\i2)[circle]{$\bx_{\i}'$} ++(180:1.5cm) node(x\i3)[circle]{$x_{\i}''$} ++(120:1.5cm) node(y\i3)[circle]{$\bx_{\i}''$};
		\foreach \x / \y in {x\i1/y\i1,y\i1/x\i2,x\i2/y\i2,y\i2/x\i3,x\i3/y\i3,y\i3/x\i1,x\i1/x\i2,x\i2/x\i3,x\i3/x\i1,y\i1/y\i2,y\i2/y\i3,y\i3/y\i1,z\i1/x\i1,z\i1/z\i3,z\i2/y\i1,z\i2/z\i3,z\i3/x\i1,z\i3/y\i1}
	 		{ \draw[-latex] (\x) -- (\y); }}
	\path (z11) ++(135:1.5cm) coordinate(r1) ++(0,1) ++(-1,0) node(d)[circle]{$d$} +(-1.5,0) node(c)[circle]{$c$} ++(1,0) ++(0,1) coordinate(r3) ++(0,.75) coordinate(ha3) ++(0,.75) coordinate(ha2) +(-1,0) node(b)[circle]{$b$} (y32) ++(300:1.5cm) ++(315:1.5cm) ++(0,-.5) coordinate(r2) node[above left]{$A_1\cup A_2$} (z32) ++(45:1.5cm) ++(0,5) coordinate(r4) ++(0,-3) node[above left]{$\{\,\ha \condition a\in A_1\cup A_2 \,\}$} ++(-1,0) coordinate(ha1) ++(0,-2) coordinate(a);
	\draw[draw opacity=0] (r2) ;
	\draw (r1) rectangle (r2);
	\draw (r3) rectangle (r4);
	\useasboundingbox;
	\path (y12) -- coordinate(helper) (x23);
	\path ($ (y12) + (0,-2) $) -| node(y1)[circle]{$y_1$} (helper);
	\path (y22) -- coordinate(helper) (x33);
	\path ($ (y22) + (0,-2) $) -| node(y2)[circle]{$y_2$} (helper);
	\draw[dashed,-latex] (a) -- (ha1);
	\draw[-latex] (b) -- (ha2);
	\draw[-latex] (c) -- (d);
	\draw[-latex] (ha3) -- (d);
	\draw[-latex] (d) .. controls +(right:1cm) and +(90:1.6cm) .. (z13);
	\draw[-latex] (d) .. controls +(right:1cm) and +(100:1.8cm) .. (z23);
	\draw[-latex] (d) .. controls +(right:1cm) and +(110:2cm) .. (z33);
	\draw[-latex] (d) .. controls +(down:6cm) and +(180:4cm) .. (y1);
	\draw[-latex] (d) .. controls +(down:9cm) and +(190:8cm) .. (y2);
	\draw[-latex] (y11) .. controls +(355:1.5cm) and +(98:3cm) .. (y1);
	\draw[-latex] (x21) .. controls +(185:1.5cm) and +(82:3cm) .. (y1);
	\draw[-latex] (x31) .. controls +(200:3.2cm) and +(0:6cm) .. (y1);
	\draw[-latex] (y21) .. controls +(355:1.5cm) and +(98:3cm) .. (y2);
	\draw[-latex] (y31) .. controls +(355:2.2cm) and +(0:5.2cm) .. (y2);
\end{tikzpicture}
\caption{Dominance graph 
  resulting from the formula $(\neg w_1 \vee w_2 \vee w_3) \wedge
  (\neg w_2 \vee \neg w_3)$ according to  Construction~\ref{cons:down-conp}.
  An edge incident to a set of alternatives represents an edge incident
  to \emph{each} alternative in the set.  The dashed edge indicates
  that $a\succ\ha$ for each $a\in A_1\cup A_2$.}
\label{fig:down-conp}
\end{figure}
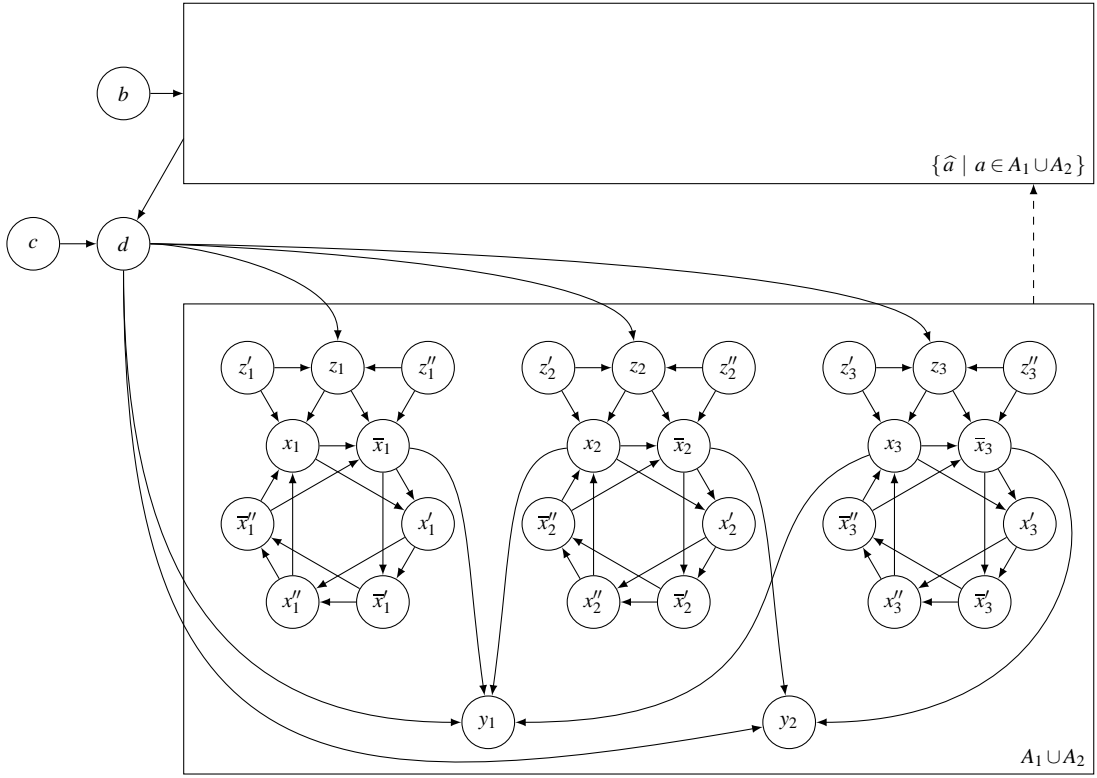

An example for this construction is shown in
Figure~\ref{fig:down-conp} for the boolean formula $(\neg w_1 \vee w_2
\vee w_3) \wedge (\neg w_2 \vee \neg w_3)$, which can be satisfied by
setting for example each of $w_1$, $w_2$, and $w_3$ to false. A
minimal downward covering set corresponding to this assignment is
$M=\{b,c\} \cup \{\overline{x}_i, \overline{x}_i', \overline{x}_i'',
z_i', z_i'' \condition 1 \leq i \leq 3\}$. Obviously, the undominated
alternatives $b$, $c$, $z_i'$, and $z_i''$, $1 \leq i \leq 3$, are
contained in every minimal downward covering set for the dominance
graph constructed. The alternative~$d$, however, is not contained in any minimal
downward covering set for~$A$. This can be seen as follows. If
$d$ were contained in some minimal downward covering set $M'$ for $A$
then none of the alternatives $\ha$ with $a \in A_1 \cup A_2$ would be
downward covered.  Hence, all
alternatives in $A_1 \cup A_2$ would necessarily
be in~$M'$, since they all dominate
a different alternative in~$M'$. But then $M'$ is no minimal downward
covering set for~$A$, since the minimal downward covering set $M$ for
$A$ is a strict subset of~$M'$.

We now show some properties of Construction~\ref{cons:down-conp} in
general.

\begin{claim}
\label{cla:down-exists}
Minimal downward covering sets are guaranteed to exist for the
dominance graph defined in Construction~\ref{cons:down-conp}.
\end{claim}

\begin{proofs}
The set $A$ of all alternatives is a downward covering set for itself.
Hence, there always exists a minimal downward covering set for the
dominance graph defined in
Construction~\ref{cons:down-conp}.~\end{proofs}

\begin{claim}
\label{cla:down-d-all}
Consider the dominance graph $(A,\succ)$
created by Construction~\ref{cons:down-conp}.
For each minimal downward covering set $M$ for $A$, if $M$ contains the
alternative $d$ then all other alternatives are contained in $M$ as
well (i.e., $A=M$).
\end{claim}

\begin{proofs}
  If $d$ is contained in some minimal downward covering set $M$ for~$A$, then
  $\{a,\ha\} \subseteq M$ for every $a \in A_1 \cup A_2$. To see this,
observe that for an arbitrary $a \in A_1 \cup A_2$ there is no $a' \in
A$ with $a' \succ \ha$ and $a' \succ d$ or with $a' \succ a$ and $a'
\succ \ha$. Since the alternatives $c$ and $b$ are undominated, they
are also in $M$, so $M=A$.~\end{proofs}

\begin{claim}
\label{cla:down-conp-satisfiable}
Consider Construction~\ref{cons:down-conp}.
The boolean formula $\varphi$ is satisfiable if and only if there is
no minimal downward covering set for $A$ that contains $d$.
\end{claim}

\begin{proofs}
For the direction from left to right, consider a satisfying assignment
$\alpha: W \rightarrow \{0,1\}$ for $\varphi$, and define the set
\[
B_{\alpha} = \{b,c\} \cup \{x_i,x_i',x_i'' \condition \alpha(w_i)=1\} \cup \{
\overline{x}_i, \overline{x}_i', \overline{x}_i'' \condition
\alpha(w_i)=0\} \cup \{z_i',z_i'' \condition 1 \leq i \leq k\}.
\]
It is not hard to verify that $B_{\alpha}$ is a minimal downward
covering set for~$A$.  Thus, there exists a minimal downward covering
set for~$A$ that
does not contain~$d$.  If there were a minimal downward covering set
$M$ for $A$ that contains~$d$, Claim~\ref{cla:down-d-all} would imply
that $M = A$.  However, since $B_{\alpha} \subset A = M$, this
contradicts minimality, so no minimal downward covering set for $A$
can contain~$d$.

For the direction from right to left, assume that no minimal downward
covering set for $A$ contains~$d$.  Since by
Claim~\ref{cla:down-exists} minimal downward covering sets are
guaranteed to exist for the dominance graph defined in
Construction~\ref{cons:down-conp}, there exists a minimal downward
covering set $B$ for $A$ that does not contain~$d$, so $B \neq A$.
It holds that
$\{z_i \condition w_i \text{ is a variable in } \varphi\}\cap B =
\emptyset$ and $\{y_j \condition f_j \text{ is a clause in } \varphi\}
\cap B = \emptyset$, for otherwise a contradiction would follow by
observing that there is no $a \in A$ with $a \succ d$ and $a \succ
z_i$, $1 \leq i \leq k$, or with $a \succ d$ and $a \succ y_j$, $1
\leq j \leq \ell$. Furthermore, we have $x_i
\not \in B$ or $\overline{x}_i \not \in B$, for each variable~$w_i \in
W$.
By external stability, for each clause $f_j$ there must exist an
alternative $a \in B$ with $a \succ y_j$. By construction and since $d
\not \in B$, we must have either $a = x_i$ for some variable $w_i$ that
occurs in $f_j$ as a positive literal, or $a=\overline{x}_i$ for some
variable $w_i$ that occurs in $f_j$ as a negative literal.  Now define
$\alpha : W \rightarrow \{0,1\}$ such that $\alpha(w_i)=1$ if $x_i \in
B$, and $\alpha(w_i)=0$ otherwise.  It is readily appreciated that
$\alpha$ is a satisfying assignment for~$\varphi$.~\end{proofs}

\begin{claim}
\label{cla:down-conp-not-satisfiable}
Consider Construction~\ref{cons:down-conp}.
The boolean formula $\varphi$ is not satisfiable if and only if there
is a unique minimal downward covering set for $A$.
\end{claim}

\begin{proofs}
We again assume that if $\varphi$ is satisfiable, it has at least two
satisfying assignments. If $\varphi$ is not satisfiable, there must be
a minimal downward covering set for $A$ that contains $d$ by
Claim~\ref{cla:down-conp-satisfiable}, and by
Claim~\ref{cla:down-d-all} there must be a minimal downward covering
set for $A$ containing all alternatives.  Hence, there is a unique minimal
downward covering set for~$A$.
Conversely, if there is a unique minimal downward covering set
for~$A$, $\varphi$ cannot be satisfiable, since otherwise there would
be at least two distinct minimal downward covering sets for~$A$,
corresponding to the distinct truth assignments for~$\varphi$, which
would yield a contradiction.~\end{proofs}

In the dominance graph created by Construction~\ref{cons:down-conp},
the minimal downward covering sets for $A$ coincide with the
minimum-size downward covering sets for~$A$. If $\varphi$ is not
satisfiable, there is only one minimal downward covering set for~$A$,
so this is also the only minimum-size downward covering set for~$A$,
and if $\varphi$ is satisfiable, the minimal downward covering sets
for $A$ correspond to the satisfying assignments of~$\varphi$.  As we
have seen in the proof of Claim~\ref{cla:down-conp-satisfiable}, these
minimal downward covering sets for $A$ always consist of $5k+2$ alternatives.
Thus, they each are also minimum-size downward covering sets for~$A$.

Merging the construction from the proof sketch of Theorem~\ref{thm:down_np}
with Construction~\ref{cons:down-conp}, we will again provide a
reduction applying
Lemma~\ref{lem:wagner-theta-2-p}, this time to downward covering set
problems.

\begin{figure}
  \centering
\begin{tikzpicture}[scale=.8]
  \scriptsize
	\tikzstyle{every circle node}=[draw,minimum size=7mm,inner sep=0pt,font=\scriptsize]
	\draw (0,0) node(d1)[circle]{$d_1$} ++(2,0) node(d2)[circle]{$d_2$} ++(2,0) node(d3)[circle]{$d_3$} ++(2,0) node(d4)[circle]{$d_4$} ++(2,0)  ++(2,0) node(d5)[circle]{$d_{2k\text{-}1}$} ++(2,0) node(d6)[circle]{$d_{2k}$};
	\foreach \i in {1,2,3,4} {
		\draw (d\i) +(130:1cm) rectangle +(1,-2) node[above left]{$A_\i$};
	}
	\draw (d5) +(130:1cm) rectangle +(1,-2) node[above left]{$A_{2k-1}$} (d6) +(130:1cm) rectangle +(1,-2) node[above left]{$A_{2k}$};
	\draw (d1) ++(60:2) node(r1)[circle]{$r_1$} (d3) ++(60:2) node(r2)[circle]{$r_2$} (d5) ++(60:2) node(r3)[circle]{$r_k$};
	\foreach \i in {1,2} {
	 \draw (r\i) +(120:2) node(s\i)[circle]{$s_\i$} +(60:2) node(t\i)[circle]{$t_\i$};
	}
	\draw (r3) +(120:2) node(s3)[circle]{$s_k$} +(60:2) node(t3)[circle]{$t_k$};
	\path (r2) -- node{$\dots$} (r3);
	\draw (t2) ++(0,2) node(d)[circle]{$d^*$} ++(150:2cm) node(c)[circle]{$c^*$};
	\foreach \x / \y in {r1/d1,r1/d2,r2/d3,r2/d4,r3/d5,r3/d6,s1/r1,s1/d1,t1/r1,t1/d2,s2/r2,s2/d3,t2/r2,t2/d4,s3/r3,s3/d5,t3/r3,t3/d6,c/d}
 		{ \draw[-latex] (\x) -- (\y); }
	\draw[-latex] (d) .. controls +(180:5cm) and +(90:3cm) .. (r1);
	\draw[-latex] (d) .. controls +(210:1cm) and +(90:2cm) .. (r2);
	\draw[-latex] (d) .. controls +(0:5cm) and +(90:3cm) .. (r3);
\end{tikzpicture}
\caption{Dominance graph from Construction~\ref{cons:down-theta}.}
 \label{fig:down-theta}
\end{figure}
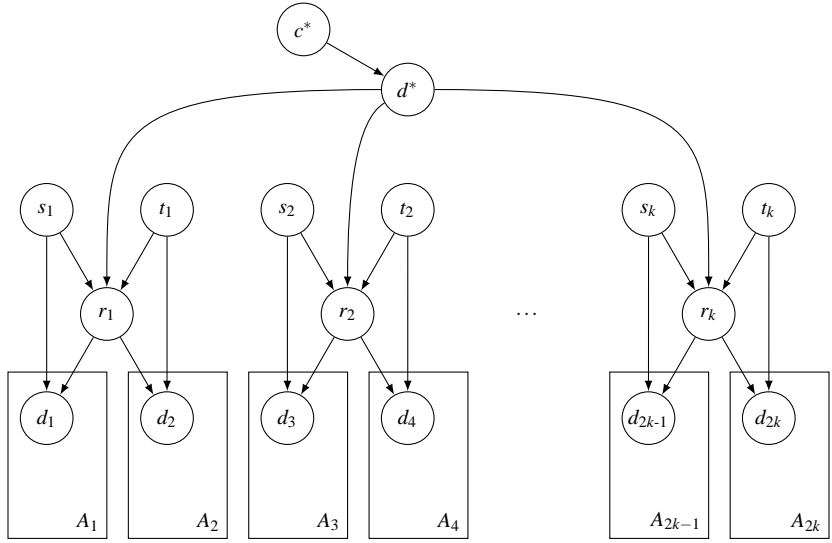

\begin{construction}[For applying Lemma~\ref{lem:wagner-theta-2-p}
to downward covering set problems]~
\label{cons:down-theta}
We again apply Wagner's Lemma with the $\np$-complete problem $S=\sat$
and construct a dominance graph. Fix an arbitrary $m \geq 1$
and let $\varphi_1,
\varphi_2, \dots , \varphi_{2m}$ be $2m$ boolean formulas 
in conjunctive normal form such that
the satisfiability of $\varphi_j$ implies the satisfiability of
$\varphi_{j-1}$, for each $j \in \{2,\dots,2m\}$. 

We will now define a polynomial-time computable function $f$, which
maps the given $2m$ boolean formulas to a dominance graph $(A,\succ)$
that has useful properties for our downward covering set problems.
The set of alternatives is
$$ A=\left(\bigcup_{i=1}^{2m} A_i\right) \cup \left( \bigcup_{i=1}^m
  \left\{r_i,s_i,t_i \right\} \right) \cup \{c^*,d^*\},$$ and the
dominance relation $\succ$ on $A$ is defined by
$$ \left(\bigcup_{i=1}^{2m} \succ_i \right) \cup 
\left( \bigcup_{i=1}^m \left\{(r_i,d_{2i-1}), (r_i,d_{2i}), (s_i,r_i),
    (s_i,d_{2i-1}), (t_i,r_i), (t_i,d_{2i}) \right\}\right) \cup
\left(\bigcup_{i=1}^k \left\{ (d^*,r_i) \right\}\right) \cup
\{(c^*,d^*)\},
$$
where we use the following notation:
\begin{enumerate}
\item For each $i$, $1 \leq i \leq m$, let $(A_{2i-1},\succ_{2i-1})$ be
  the dominance graph that results from the formula $\varphi_{2i-1}$
  according to Brandt and Fischer's construction given in the proof
  sketch of Theorem~\ref{thm:down_np}. We will again use the same
  names for the alternatives in $A_{2i-1}$ as in that proof sketch,
  except that we attach the subscript $2i-1$.
\item For each $i$, $1 \leq i \leq m$, let $(A_{2i},\succ_{2i})$ be the
  dominance graph that results from the formula $\varphi_{2i}$
  according to Construction~\ref{cons:down-conp}. We will again use
  the same names for the alternatives in $A_{2i}$ as in that
  construction, except that we attach the subscript $2i$.
\item For each $i$, $1 \leq i \leq m$, the dominace graphs
  $(A_{2i-1},\succ_{2i-1})$ and $(A_{2i},\succ_{2i})$ are connected by
  the alternatives $s_i$, $t_i$, and $r_i$ (which play a similar role
  as the alternatives $z_i$, $z_i'$, and $z_i''$ for each variable in
  Construction~\ref{cons:down-conp}). The resulting dominance graph is
  denoted by $(B_i,\succ_i^B)$.
\item Connect the $m$ dominance graphs $(B_i,\succ_i^B)$, $1\leq i
  \leq m$ (again similarly as in
  Construction~\ref{cons:down-conp}). The alternative $c^*$ dominates
  $d^*$, and $d^*$ dominates the $m$ alternatives $r_i$, $ 1\leq i
  \leq m$.
\end{enumerate}
\end{construction}
This construction is illustrated in
Figure~\ref{fig:down-theta}. Clearly, $(A,\succ)$ is computable in
polynomial time.

\begin{claim}
\label{cla:down-each-M_i-is-included}
Consider Construction~\ref{cons:down-theta}.
For each $i$, $1 \leq i \leq 2m$, let $M_i$ be the minimal downward
covering set for $(A_i,\succ_i)$. Then each of the sets $M_i$ must
be contained in every minimal downward covering set for $(A,\succ)$.
\end{claim}

\begin{proofs}
For each~$i$, $1 \leq i \leq 2m$, the only
alternative in $A_i$ dominated from outside $A_i$ is~$d_i$.  Since
$d_i$ is also dominated by the undominated alternative $z_{1,i} \in
A_i$ for odd~$i$, and by the undominated alternative $c_i \in A_i$ for
even~$i$, it is readily appreciated that internal and external
stability with respect to elements of $A_i$ only depends on the
restriction of the dominance graph to~$A_i$.~\end{proofs}

\begin{claim}
\label{cla:wagner-mdcm}
Consider Construction~\ref{cons:down-theta}.  It holds that
\begin{eqnarray}
\label{eqn:wagner-mdcm}
\lefteqn{\|\{i\condition \varphi_i \in \sat\}\| \text{ is odd}}
\nonumber \\ 
& \Longleftrightarrow & 
\mbox{$d^*$ is contained in some minimal downward covering set $M$ for $A$}.
\end{eqnarray}
\end{claim}

\begin{proofs}
For the direction from left to right in~(\ref{eqn:wagner-mdcm}),
assume that $\|\{i \condition
\varphi_i \in \sat \} \|$ is odd.  Thus, there is
some $j \in \{1,\dots,m\}$ such that $\varphi_1, \varphi_2, \ldots,
\varphi_{2j-1}$ are each satisfiable and
$\varphi_{2j}, \varphi_{2j+1}, \ldots, \varphi_{2m}$ are each not. Define
$$M= \left(\bigcup_{i=1}^{2m} M_i \right) 
\cup \left(\bigcup_{i=1}^m \left\{s_i,t_i \right\} \right) \cup
\left\{ r_j,c^*,d^* \right\}, $$ where for each $i$, $1\leq i \leq
2m$, $M_i$ is some minimal downward covering set of the restriction of
the dominance graph to $A_i$, satisfying that $d_i \in M_i$ if and only if
\begin{enumerate}
\item $i$ is odd and $\varphi_i$ is satisfiable, or 
\item $i$ is even and $\varphi_i$ is not satisfiable.
\end{enumerate}
Such sets $M_i$ exist by the proof sketch of Theorem~\ref{thm:down_np} and by
Claim~\ref{cla:down-conp-satisfiable}. In particular, $\varphi_{2j-1}$
is satisfiable and $\varphi_{2j}$ is not, so $\{d_{2j-1},d_{2j}\}
\subseteq M$. There is no alternative that dominates $d_{2j-1}$,
$d_{2j}$, and $r_j$.  Thus, $r_j$ must be in~$M$.  The other
alternatives $r_i$, $1\leq i \leq m$ and $i\neq j$, are 
downward covered by either $s_i$ if 
$d_{2i-i} \not \in M$, or $t_i$ if $d_{2i} \not \in M$.
Finally, $d^*$ cannot be downward covered, because $d^* \succ r_j$ and
no alternative dominates both $d^*$ and~$r_j$.  Internal and external
stability with respect to the elements of~$M_i$, as well as minimality of
$\bigcup_{i=1}^{2k} M_i$, follow from the proofs of
Theorem~\ref{thm:down_np} and Claim~\ref{cla:down-conp-satisfiable}.
All other elements of $M$ are undominated and thus contained in every
downward covering set. We conclude that $M$ is a minimal downward
covering set for $A$ that contains~$d^*$.

For the direction from right to left in~(\ref{eqn:wagner-mdcm}),
assume that there exists a
minimal downward covering set $M$ for $A$ with $d^* \in M$.
By internal stability, there must exist some $j$, $1 \leq j \leq k$, such
that $r_j \in M$.  Thus, $d_{2j-1}$ and $d_{2j}$ must be in~$M$,
too. It then follows from the proof sketch of Theorem~\ref{thm:down_np} and
Claim~\ref{cla:down-conp-satisfiable} that $\varphi_{2j-1}$ is
satisfiable and $\varphi_{2j}$ is not.  Hence, $\|\{i \condition
\varphi_i \in \sat \} \|$ is odd.~\end{proofs}

By the remark made after Theorem~\ref{thm:down_np}, 
Construction~\ref{cons:down-theta}
cannot be used straightforwardly to obtain complexity
results for minimum-size downward covering sets.

\section{Proof of Theorem~\ref{thm:results}}
\label{sec:proofs}

In this section, we prove Theorem~\ref{thm:results} by applying the
constructions and the properties of the resulting dominance graphs
presented in Section~\ref{sec:constructions}.  We start with the
results on minimal and minimum-size upward covering sets.

\subsection{Minimal and Minimum-Size Upward Covering Sets}

\begin{theorem}
\label{thm:exist-bounded}
It is $\np$-complete to decide, given a dominance graph $(A,\succ)$
and a positive integer~$k$, whether there exists a
minimal/minimum-size upward covering set for $A$ of size at most~$k$.
That is, both $\mucexists$ and $\msucexists$ are $\np$-complete.
\end{theorem}
\begin{proofs}
  This result can be proven by using the construction of
  Theorem~\ref{thm:up_np}. Let $\varphi$ be a given boolean formula 
  in conjunctive normal form,
  and let $n$ be the number of variables occuring in~$\varphi$.
  Setting the bound $k$ for the size of a minimal/minimum-size upward
  covering set to $2n+1$ proves that both problems are hard for~$\np$.
  Indeed,
  as we have seen in the paragraph after the proof sketch of
  Theorem~\ref{thm:up_np}, there is a size $2n+1$ minimal upward
  covering set (and hence a minimum-size upward covering set) for $A$
  if and only if $\varphi$ is satisfiable.
  Both problems are $\np$-complete, since they can obviously be
  decided in nondeterministic polynomial time.~\end{proofs}

\begin{theorem}
\label{thm:minimal-in-some}
Deciding whether a designated alternative is contained in some minimal
upward covering set for a given dominance graph 
is hard for $\Theta_2^p$ and in~$\Sigma_2^p$.
That is, $\mucmember$ is hard for $\Theta_2^p$ and in~$\Sigma_2^p$.
\end{theorem}
\begin{proofs}
  $\Theta_2^p$-hardness follows directly from
  Claim~\ref{cla:wagner-mucm}.  For the upper bound, let $(A,\succ)$ be a
  dominance graph and $d$ a designated alternative in $A$. First,
  observe that we can verify in polynomial time whether a subset of
  $A$ is an upward covering set for~$A$, simply by checking whether it
  satisfies internal and external stability. Now, we can guess an
  upward covering set $B\subseteq A$ with $d\in B$ in nondeterministic
  polynomial time and verify its minimality by checking that none of
  its subsets is an upward covering set for~$A$.
  This places the problem in
  $\np^{\conp}$ and consequently in~$\Sigma_2^p$.~\end{proofs}

\begin{theorem}
\label{thm:minimumsize-in-some}
\begin{enumerate}
\item It is $\Theta_2^p$-complete to decide whether a designated
  alternative is contained in some minimum-size upward covering set
  for a given dominance graph.
  That is, $\msucmember$ is $\Theta_2^p$-complete.
\item It is $\Theta_2^p$-complete to decide whether a designated
  alternative is contained in all minimum-size upward covering sets
  for a given dominance graph.
  That is, $\msucmemberall$ is $\Theta_2^p$-complete.
\end{enumerate}
\end{theorem}

\begin{proofs}
  By the remark made after Claim~\ref{cla:wagner-mucm}, both
  problems are hard for $\Theta_2^p$.

  To see that $\msucmember$ is contained in~$\Theta_2^p$, let
  $(A,\succ)$ be a dominance graph and $d$ a designated alternative in~$A$.
  Obviously, in nondeterministic polynomial time we can decide, given
  $(A,\succ)$, $x \in A$, and some positive integer $\ell \leq \|A\|$,
  whether there exists some upward covering set $B$ for $A$ such that
  $\|B\| \leq \ell$ and $x \in B$. Using this problem as an $\np$ oracle,
  in~$\Theta_2^p$ we can decide, given $(A,\succ)$ and $d \in A$,
  whether there exists a minimum-size upward covering set for $A$
  containing $d$ as follows. The oracle is asked whether for each pair
  $(x,\ell)$, where $x \in A$ and $1 \leq \ell \leq \|A\|$, there
  exists an upward covering set for $A$ of size bounded by $\ell$ that
  contains the alternative~$x$. The number of queries is polynomial
  (i.e., in $\mathcal{O}(\|A\|^2)$),
and all queries can be asked in parallel. Having all the answers,
determine the size $k$ of a minimum-size upward covering set for~$A$,
and accept if the oracle answer to $(d,k)$ was yes, otherwise reject.

To show that $\msucmemberall$ is in $\Theta_2^p$, let $(A,\succ)$ be a
dominance graph and $d$ a designated alternative in~$A$.  We now use
as our oracle the set of all $((A,\succ), x, \ell)$, where $(A,\succ)$
is our dominance graph, $x \in A$ is an alternative, and $\ell \leq
\|A\|$ a positive integer, such that there exists some upward covering
set $B$ for $A$ with $\|B\| \leq \ell$ and $x \not\in B$.  Clearly,
this problem is also in~$\np$, and the size $k$ of a minimum-size upward
covering set for $A$ can again be determined by asking
$\mathcal{O}(\|A\|^2)$ queries in parallel (if all oracle answers are
no, it holds that $k=\|A\|$).  Now, the $\Theta_2^p$
machine accepts its input $((A,\succ),d)$ if the oracle answer for the
pair $(d,k)$ is no, and otherwise it rejects.~\end{proofs}

\begin{theorem}
\label{thm:minimal-all-unique-is}
\begin{enumerate}
\item (Brandt and Fischer~\cite{bra-fis:j:minimal-covering-set})
  It is $\conp$-complete to decide whether a designated
  alternative is contained in all minimal upward covering sets for a
  given dominance graph.  That is, $\mucmemberall$ is $\conp$-complete.
\item It is $\conp$-complete to decide whether a given subset of the
  alternatives is a
  minimal upward covering set for a given dominance graph.
  That is, $\muctest$ is $\conp$-complete.
\item It is $\conp$-hard and in $\Sigma_2^p$ to decide whether there is a unique
  minimal upward covering set for a given dominance graph.
  That is, $\mucunique$ is $\conp$-hard
and in $\Sigma_2^p$.
\end{enumerate}
\end{theorem}

\begin{proofs}
  It follows from Claim~\ref{cla:notsatisfiable-all} that $\varphi$ is
  not satisfiable if and only if the entire set of alternatives $A$ is
  a (unique) minimal upward covering set for~$A$. Furthermore, if
  $\varphi$ is satisfiable, there exists more than one minimal upward
  covering set for~$A$ and none of them contains~$e_1$ (provided that
  $\varphi$ has more than one satisfying assignment, which can be
  ensured, if needed, by adding a dummy variable such that the 
  satisfiability of
  the formula is not affected).
  This proves $\conp$-hardness for all three problems.
  $\mucmemberall$ and $\muctest$ are also
  \emph{contained} in $\conp$, as they can be decided in the
  positive by checking whether there does \emph{not} exist an upward
  covering set that satisfies certain properties related to the
  problem at hand, so they both are
  $\conp$-complete. $\mucunique$ can be decided in the positive by
  checking whether there exists an upward covering set $M$ such that
  all sets that are not strict supersets of $M$ are \emph{not} upward
  covering sets for the set of all alternatives. Thus, $\mucunique$ is
  in $\Sigma_2^p$.~\end{proofs}

The first statement of Theorem~\ref{thm:minimal-all-unique-is}
was already shown by Brandt and
Fischer~\cite{bra-fis:j:minimal-covering-set}.  However, their
proof---which uses essentially the reduction from the proof of
\thmref{thm:up_np}, except that they start from the $\conp$-complete
problem {\sc Validity} (which asks whether a given formula is valid,
i.e., 
true under every
assignment~\cite{pap:b-1994:complexity})---does not yield any of the
other $\conp$-hardness results in Theorem~\ref{thm:minimal-all-unique-is}.

\begin{theorem}
\label{thm:minimumsize-is}
It is $\conp$-complete to decide whether a given subset of the
alternatives is a minimum-size upward covering set for a given
dominance graph.  That is, $\msuctest$ is $\conp$-complete.
\end{theorem}

\begin{proofs}
  This problem is in $\conp$, since it can be decided in the positive
  by checking whether the given subset $M$ of alternatives is an
  upward covering set for the set $A$ of all alternatives (which is
  easy) and all sets of smaller size than $M$ are not upward covering
  sets for~$A$ (which is a $\conp$ predicate),  and $\conp$-hardness
  follows directly from Claim~\ref{cla:notsatisfiable-all}.~\end{proofs}

\begin{theorem}
\label{thm:minimumsize-unique}
Deciding whether there exists a unique minimum-size upward
covering set for a given dominance graph is hard for $\conp$ and
in~$\Theta_2^p$.  That is, $\msucunique$ is $\conp$-hard and
in~$\Theta_2^p$.
\end{theorem}

\begin{proofs}
It is easy to see that
$\conp$-hardness follows directly from the $\conp$-hardness of
$\mucunique$  (see
Theorem~\ref{thm:minimal-all-unique-is}). Membership in $\Theta_2^p$
can be proven by using the same oracle as in the proof of the first
part of Theorem~\ref{thm:minimumsize-in-some}. We ask for all pairs
$(x,\ell)$, where $x \in A$ and $1 \leq \ell \leq \|A\|$, whether
there is an upward
covering set $B$ for $A$ such that $\|B\| \leq \ell$ and $x \in B$.
Having all the
answers, determine the minimum size $k$ of a minimum-size upward
covering set for $A$. Accept if there are exactly $k$ distinct alternatives
$x_1,\ldots , x_k$ for which the answer for $(x_i,k)$, $1\leq i \leq k$,
was yes, otherwise reject.~\end{proofs}

An important consequence of the proofs of
Theorems~\ref{thm:minimal-all-unique-is}
and~\ref{thm:minimumsize-unique} (and of \consref{cons:conp} that
underpins these proofs) regards the hardness of
the search problems $\mucfind$ and $\msucfind$.

\begin{theorem}\label{thm:search}
Assuming $\p \neq \np$, neither minimal upward covering sets nor
minimum-size upward covering sets can be found in polynomial
time. That is, neither $\mucfind$ nor $\msucfind$ are polynomial-time
computable unless $\p= \np$.
\end{theorem}

\begin{proofs}
Consider the problem of deciding whether
there exists a \emph{nontrivial} minimal/minimum-size upward covering set, \ie
one that does \emph{not} contain all
alternatives.
By
Construction~\ref{cons:conp}
that is
applied in proving
Theorems~\ref{thm:minimal-all-unique-is}
and~\ref{thm:minimumsize-unique}, there exists a trivial
minimal/minimum-size upward covering set for $A$ (i.e.,
one containing all alternatives in~$A$) if and only if this set is the only
minimal/minimum-size upward covering set for $A$.  Thus, the
$\conp$-hardness
proof for the problem of deciding whether there is a unique
minimal/minimum-size
upward covering set for $A$ (see the proofs of
Theorem~\ref{thm:minimal-all-unique-is} and~\ref{thm:minimumsize-unique})
immediately implies that the problem of deciding whether there is a
nontrivial minimal/minimum-size upward covering set for $A$ is $\np$-hard.
However, since
the latter problem can easily be reduced to the
search problem (because the search problem,
when used as a function oracle,
will yield the set of all alternatives if and only if
this set is the only minimal/minimum-size upward covering set for~$A$),
it follows
that the search problem cannot be solved in polynomial time
unless $\p=\np$.~\end{proofs}

\subsection{Minimal and Minimum-Size Downward Covering Sets}

\begin{theorem}
\label{thm:down-exist-bounded}
It is $\np$-complete to decide, given a dominance graph $(A,\succ)$
and a positive integer~$k$,  whether there exists a
minimal/minimum-size downward covering set for $A$ of size at most~$k$.
That is, $\mdcexists$ and $\msdcexists$ are both $\np$-complete.
\end{theorem}

\begin{proofs}
  Membership in $\np$ is obvious, since we can nondeterministically
  guess a subset $M \subseteq A$
  of the alternatives with $\|M\| \leq k$ and can then
  check in polynomial time whether $M$
  is a downward covering set for~$A$.
  $\np$-hardness of $\mdcexists$ and
  $\msdcexists$ follows from \consref{cons:down-conp}, the proof of
  Claim~\ref{cla:down-conp-satisfiable}, and the comments made after
  Claim~\ref{cla:down-conp-not-satisfiable}: If $\varphi$ is a
  given formula with $n$ variables, then there exists a 
  minimal/minimum-size downward
  covering set of size $5n+2$ if and only if $\varphi$ is
  satisfiable.~\end{proofs}

\begin{theorem}
\label{thm:down-minimumsize-insome-all-unique}
$\msdcmember$, $\msdcmemberall$, and $\msdcunique$ are $\conp$-hard
and in~$\Theta_2^p$.
\end{theorem}

\begin{proofs}
It follows from Claim~\ref{cla:down-conp-not-satisfiable} that
$\varphi$ is not satisfiable if and only if the entire set $A$ of all
alternatives is the unique minimum-size downward covering set for
itself.  Moreover, assuming that $\varphi$ has at least two satisfying
assignments, if $\varphi$ is satisfiable, there are at least two
distinct minimum-size downward covering sets for~$A$.  This shows that
each of $\msdcmember$, $\msdcmemberall$, and $\msdcunique$ is
$\conp$-hard.  For all three problems, membership in $\Theta_2^p$ is
shown 
similarly to the proofs of the corresponding minimum-size
upward covering set problems. However, since downward covering sets may
fail to exist, the proofs must be slightly adapted. For $\msdcmember$
and $\msdcunique$, the machine rejects the input if the size $k$ of a
mininum-size downward covering set cannot be computed (simply because
there doesn't exist any such set). For
$\msdcmemberall$, if all oracle answers are no, it must be checked
whether the set of all alternatives is a downward covering set for
itself. If so, the machine accepts the input, otherwise it
rejects.~\end{proofs}

\begin{theorem}
\label{thm:down-minimumsize-is}
It is $\conp$-complete to decide whether a given subset is a
minimum-size downward
covering set for a given dominance graph.
That is, $\msdctest$ is $\conp$-complete.
\end{theorem}

\begin{proofs}
This problem is in $\conp$, since its complement (i.e., the problem of
deciding whether a given subset of the set $A$ of alternatives is not
a minimum-size downward covering set for~$A$) can be decided in
nondeterministic polynomial time. Hardness for $\conp$ follows
directly from Claim~\ref{cla:down-conp-not-satisfiable}.~\end{proofs}

\begin{theorem}
\label{thm:down-minimumsize-in-some}
Deciding whether a designated alternative is contained in some minimal
downward covering set for a given dominance graph is hard for
$\Theta_2^p$ and in $\Sigma_2^p$.
That is, $\mdcmember$ is hard for $\Theta_2^p$ and in~$\Sigma_2^p$.
\end{theorem}

\begin{proofs}
  Membership in $\Sigma_2^p$ can be shown analogously to the proof of
  Theorem~\ref{thm:minimal-in-some}, and $\Theta_2^p$-hardness follows
  directly from Claim~\ref{cla:wagner-mdcm}.~\end{proofs}

\begin{theorem}
\label{thm:down-minimal-all-unique-is}
\begin{enumerate}
\item (Brandt and Fischer~\cite{bra-fis:j:minimal-covering-set}) It is $\conp$-complete to decide whether a designated alternative is
contained in all minimal downward covering sets for a given dominance
graph.  That is, $\mdcmemberall$ is $\conp$-complete.
\item It is $\conp$-complete to decide whether a given subset of the
  alternatives is a
  minimal downward covering set for a given dominance graph.
  That is, $\mdctest$ is $\conp$-complete.
\item It is $\conp$-hard and in $\Sigma_2^p$ to decide whether there is a unique minimal
  downward covering set for a given dominance graph.
  That is, $\mdcunique$ is $\conp$-hard
and in $\Sigma_2^p$.
\end{enumerate}
\end{theorem}

\begin{proofs}
It follows from Claim~\ref{cla:down-conp-not-satisfiable} that
$\varphi$ is not satisfiable if and only if the entire set of
alternatives $A$ is a unique minimal downward covering set for
$A$. Furthermore, if $\varphi$ is satisfiable, there exists more than
one minimal downward covering set for $A$ and none of them contains $d$
(provided that $\varphi$ has more than one satisfying assignment,
which can be ensured, if needed, by adding a dummy variable such that the
satisfiability of the formula is not affected).
This proves $\conp$-hardness for all three problems.
$\mdcmemberall$ and $\mdctest$ are also contained in $\conp$, because
they can be
decided in the positive by checking whether there does not exist a
downward covering set that satisfies certain properties related to the
problem at hand. Thus, they are
both $\conp$-complete. $\mdcunique$ can be decided in the positive by
checking whether there exists a downward covering set $M$ such that
all sets that are not strict supersets of $M$ are \emph{not} downward
covering sets for the set of all alternatives. This shows that
$\mdcunique$ is in $\Sigma_2^p$.~\end{proofs}

The first statement of Theorem~\ref{thm:down-minimal-all-unique-is}
was already shown by Brandt
and Fischer~\cite{bra-fis:j:minimal-covering-set}. However, their
proof---which uses essentially the reduction from the proof of
Theorem~\ref{thm:down_np}, except that they start from the
$\conp$-complete problem {\sc Validity}---does 
not yield any of the other
$\conp$-hardness results in Theorem~\ref{thm:down-minimal-all-unique-is}.

An important consequence of the proofs of
Theorems~\ref{thm:down-minimumsize-insome-all-unique}
and~\ref{thm:down-minimal-all-unique-is}
regards the hardness of
the search problems $\mdcfind$ and $\msdcfind$.
(Note that the hardness of $\mdcfind$ also follows from
a result by Brandt and Fischer~\cite[Thm.~9]{bra-fis:j:minimal-covering-set},
see the discussion in Section~\ref{sec:results}.)

\begin{theorem}\label{thm:down-search}
Assuming $\p \neq \np$, neither minimal downward covering sets nor
minimum-size downward covering sets can be found in polynomial time
(i.e., neither $\mdcfind$ nor $\msdcfind$ are polynomial-time
computable unless $\p=\np$),
even when the existence of a downward covering set is guaranteed.
\end{theorem}

\begin{proofs}
Consider the problem of deciding whether
there exists a \emph{nontrivial} minimal/minimum-size downward covering set,
\ie 
one that does \emph{not} contain all
alternatives.
By
Construction~\ref{cons:down-conp}
that is
applied in proving
Theorems~\ref{thm:down-minimumsize-insome-all-unique}
and~\ref{thm:down-minimal-all-unique-is}, there exists a
trivial
minimal/minimum-size downward covering set for $A$ (i.e.,
one containing all alternatives in~$A$) if and only if this set is the only
minimal/minimum-size downward covering set for~$A$.  Thus, the
$\conp$-hardness
proof for the problem of deciding whether there is a unique
minimal/minimum-size
downward covering set for $A$ (see the proofs of
Theorems~\ref{thm:down-minimumsize-insome-all-unique}
and~\ref{thm:down-minimal-all-unique-is})
immediately implies that the problem of deciding whether there is a
nontrivial minimal/minimum-size downward covering set for $A$ is $\np$-hard.
However, since
the latter problem can easily be reduced to the
search problem (because the search problem,
when used as a function oracle,
will yield the set of all alternatives if and only if
this set is the only minimal/minimum-size downward covering set for
$A$), it follows
that the search problem cannot be solved in polynomial time
unless $\p=\np$.~\end{proofs}

\bibliography{covering}
\bibliographystyle{alpha}

\end{document}